\documentstyle[12pt]{article}
\renewcommand{\thesection}{\Roman{section}}
\renewcommand{\theequation}{\arabic{section}.\arabic{equation}}
\newcommand{\be}{\begin{equation}}
\newcommand{\ee}{\end{equation}}

\def\b{\bigskip}

\def\mybox{\sqcap\kern-.66em\sqcup\kern.66em}

\def\no{\noindent}

\def\ve{\vfill\eject}

\def\e e{$e^+ e^-$ }

\begin{document}
\renewcommand{\thefootnote}{\fnsymbol{footnote}}
\begin{flushright}
UCLA/94/TEP/8
\end{flushright}
\begin{flushright}
February 1994
\end{flushright}
\begin{center}{\Large\bf Structure of the Chiral Scalar Superfield \\
in Ten Dimensions}
\end{center}
\vskip .5in
\begin{center}
{\large P. S. Kwon} \\
Department of Physics, Kyungsung University, Pusan 608-736, Korea \\
\vskip .1in
and\\
\vskip .1in
{\large M. Villasante}\footnote
{Work supported in part by the World Laboratory.}\\
Department of Physics, University of California, Los Angeles, CA 90024-1547
\end{center}
\vskip 1in
\begin{center}{\bf\large Abstract}
\end{center}
\vskip 12pt

        We describe the tensors and spinor-tensors included in the
$\theta$-expansion of the ten-dimensional chiral scalar superfield. The
product decompositions of all the irreducible structures with $\theta$ and
the $\theta^2$ tensor are provided as a first step towards the obtention of
a full tensor calculus for the superfield.

\thispagestyle{empty}
\addtocounter{page}{-1}
\newpage

\section{Introduction}

The field structure of higher dimensional supergravities as well as
of $N\geq 3$ extended supergravities is still an open problem.
It is an old problem whose general solution
was deemed impossible for a while due to some ``no-go theorems"  \cite{Taylor}
establishing the impossibility of writing quadratic Lagrangians for the
linearized (free) theory. The underlying problem was the so-called
``self-duality counting paradox" \cite{Roc} which was subsequently resolved
\cite{Siegel} by the discovery of the fact that the Lagrangian for the
linear theory is not quadratic when is dealing with fields having a self-dual
field strength.

In particular one would really like to know the auxiliary field structure of
10-dimensional supergravity \cite{diezd}, a theory unaffected by the
above mentioned no-go theorems, due to its relevance for string theory
applications.

Traditionally the auxiliary field structures for supergravities that are known
have always been found in a rather {\it ad hoc} manner by counting degrees of
freedom and trying to add suitable new fields in order to match the bosonic and
fermionic degrees of freedom off-shell \cite{std}.  It was only later, after
the answer was known, that more systematic ways of deriving the result were
found.  However, for the more complicated theories the auxiliary field
structure
becomes so complex that it has been impossible to guess.  Complicating matters
further is the above-mentioned self-duality counting paradox, and we are
finally bound to use a systematic approach to solve the problem.

A fruitful approach in 4 dimensions is the use of the superconformal framework
in which the different Poincar\'e supergravities correspond to using different
compensators to fix the extra degree of freedom \cite{Kugo}.  However, while
the
super-Poincar\'e algebra remains essentially the same in higher dimensions, the
same is not true for the superconformal one which acquires a multitude of new
generators \cite{vanHolten}, which complicates enormously this gauge-fixing
procedure. In fact, even though the complete off-shell structure of
ten-dimensional conformal supergravity was obtained long ago in \cite{dewit1},
a satisfactory off-shell Poincar\'{e} version is still lacking (see
\cite{dewit2,howe}).

In ref. \cite{howe} it was proposed a linearized off-shell 10-dimensional
supergravity adding to the conformal supergravity multiplet a set of 2
full-fledged chiral scalar superfields. However
this is in all likelihood a reducible
version since each chiral scalar superfield contains 3 irreducible pieces
\cite{Man2}. Furthermore, the tensorial structure and transformation rules
of the component fields was not provided, even at the linearized level.

A second more promising approach is the irreducible superfield method, which
has been successfully used in the $N=1$ \cite{sokatchev} and $N=2$ \cite{kim}
cases.  In working with superfields \cite{salam}
one is automatically assured that the numbers of
fermionic and bosonic degrees of freedom will match, but general superfields
are usually objects too large to handle, containing many more fields that one
is interested in, especially in higher dimensions (though some interesting
four-dimensional results have been obtained using unconstrained superfields
in the so called harmonic superspace approach \cite{harm}).
That is why the importance
of irreducible superfields, which are much simpler objects
satisfying additional supersymmetric constraints.  These subsidiary conditions
are usually differential equations involving the superspace covariant
derivatives, and can be obtained by applying appropriate projection operators
for the corresponding eigenvalues of the Casimirs \cite{sokatchev}.
The Casimir operators for the super-Poincar\'e algebras in all dimensions
are known and they have been used to decompose the 11-dimensional \cite{Man1}
and 10-dimensional massive scalar superfields.  In the 10-dimensional case,
there is an additional interesting complication, namely that the lowest
(quadratic) Casimir operator $C_2$ does not distinguish between the
3 irreducible pieces since it has the same eigenvalue for the corresponding
representation \cite{Man2}.  Therefore one would have to construct projection
operators using the second lowest (quartic) Casimir operator $C_4$, which
does distinguish among those
representations, but the resulting differential equations are so complicated
as to render the method impractical.  However, this difficulty was circumvented
by resorting to the Cartan subalgebra in order to obtain simple differential
equations which were used to characterize the irreducible pieces of the
massless and massive 10-dimensional scalar superfield in \cite{kwon1} and
\cite{kwon2} respectively. The irreducible superfields were then obtained as
expansions in Grassmann-Hermite polynomials, but the field components in these
non-covariant expressions remained to be sorted out, though in principle
it can be done.

In all this one final basic stumbling block remains though:
while it is known from
group theory methods what are the fields contained in scalar superfield
\cite{Man3}, it is
not known in what form they appear.  In other words, while it is trivial to
write the scalar superfield in multispinor language:

$$\Phi(x,\theta)=\sum_{j=0}^{16} \chi_{\alpha_1\ldots \alpha_j}(x)
  \theta^{\alpha_1}\ldots\theta^{\alpha_j},\eqno(1.1)$$

\no it is a rather different proposition to extract the irreducible fields with
their tensor (non-spinor) indices out of the $\chi_{\alpha_1\ldots\alpha_j}(x)$
fields.  The latter is equivalent to decompose into irreducible pieces all the
possible powers of the anticommuting variable $\theta^\alpha$, and that is what
we will do in this paper.
The irreducible SO(10) representations contained in the corresponding powers of
$\theta$ are reproduced in Table 1. The list is for increasing powers of one
of the basic spinorial representations
$[\frac{1}{2} \frac{1}{2} \frac{1}{2} \frac{1}{2} \frac{1}{2}]$ corresponding
to the positive chirality projection $\theta^{(+)}$. For the negative chirality
case $\theta^{(-)}$ one just needs to read Table 1 upside down. In either case
the representations corresponding to the fields
$\chi_{\alpha_1\ldots \alpha_j}(x)$ are the same but with opposite chirality
and duality when they apply. In other words, the representations for the
fields accompanying a certain power of $\theta^{(+)}$ are given by the same
power of $\theta^{(-)}$ and viceversa.

\begin{table}
\centering
\begin{tabular}{|c|c|c|}   \hline
$j$ & $\theta^{\alpha_1}\ldots\theta^{\alpha_j}$ & Dimension \\ \hline
$0$ & $[0]$ & $1$ \rule{0cm}{0.5cm}\\
$1$ & $[\frac{1}{2} \frac{1}{2} \frac{1}{2} \frac{1}{2} \frac{1}{2}]$ &
$16$ \rule{0cm}{0.5cm}\\
$2$ & $[1 \; 1 \; 1]$ & $120$ \rule{0cm}{0.5cm} \\
$3$ & $[\frac{3}{2} \frac{3}{2} \frac{1}{2} \frac{1}{2} \frac{-1}{2}]$ &
$560$ \rule{0cm}{0.5cm} \\
$4$ & $[2 \;  2] \oplus [2 \;  1 \;  1 \;  1 -1]$ & $770 + 1050$
\rule{0cm}{0.5cm} \\
$5$ & $[\frac{5}{2} \frac{3}{2} \frac{1}{2} \frac{1}{2} \frac{-1}{2}] \oplus
    [\frac{3}{2} \frac{3}{2} \frac{3}{2} \frac{3}{2} \frac{-3}{2}]$ &
$3696 + 672$ \rule{0cm}{0.5cm} \\
$6$ & $[3 \; 1 \; 1] \oplus [2 \; 2 \; 1 \; 1 -1]$ & $4312 + 3696$
\rule{0cm}{0.5cm} \\
$7$ & $[\frac{7}{2} \frac{1}{2} \frac{1}{2} \frac{1}{2} \frac{1}{2}] \oplus
    [\frac{5}{2} \frac{3}{2} \frac{3}{2} \frac{1}{2} \frac{-1}{2}]$ &
$2640 + 8800$ \rule{0cm}{0.5cm} \\
$8$ & $[4] \oplus [3 \; 1 \; 1 \; 1] \oplus [2 \; 2 \; 2]$ & $660 + 8085
+ 4125$ \rule{0cm}{0.5cm} \\
$9$ & $[\frac{7}{2} \frac{1}{2} \frac{1}{2} \frac{1}{2} \frac{-1}{2}] \oplus
    [\frac{5}{2} \frac{3}{2} \frac{3}{2} \frac{1}{2} \frac{1}{2}]$ &
$2640 + 8800$ \rule{0cm}{0.5cm} \\
$10$ & $[3 \; 1 \; 1] \oplus [2 \; 2 \; 1 \; 1 \; 1]$ & $4312 + 3696$
\rule{0cm}{0.5cm} \\
$11$ & $[\frac{5}{2} \frac{3}{2} \frac{1}{2} \frac{1}{2} \frac{1}{2}] \oplus
    [\frac{3}{2} \frac{3}{2} \frac{3}{2} \frac{3}{2} \frac{3}{2}]$ &
$3696 + 672$ \rule{0cm}{0.5cm} \\
$12$ & $[2 \; 2] \oplus [2 \; 1 \; 1 \; 1 \; 1]$ & $770 + 1050$
\rule{0cm}{0.5cm} \\
$13$ & $[\frac{3}{2} \frac{3}{2} \frac{1}{2} \frac{1}{2} \frac{1}{2}]$ &
$560$ \rule{0cm}{0.5cm}\\
$14$ & $[1 \; 1 \; 1]$ & $120$ \rule{0cm}{0.5cm}\\
$15$ & $[\frac{1}{2} \frac{1}{2} \frac{1}{2} \frac{1}{2} - \frac{1}{2}]$ &
$16$ \rule{0cm}{0.5cm}\\
$16$ & $[0]$ & $1$ \rule[-.5cm]{0cm}{1cm}\\ \hline
\end{tabular}
\caption{Decomposition of the totally antisymmetrized Kronecker (wedge)
powers of the basic spinor representation of SO(10), as given by their
highest weights.}
\end{table}

\ve

\section{Fierz Identity}
\setcounter{equation}{0}

\b

The 10-dimensional Fierz identity for strictly anticommuting $\theta$'s can be
put in a very simple form

\be
\bar\theta^{(\pm)} O_1\theta^{(\pm)}\bar\theta^{(\pm)}O_2\theta^{(\pm)}
  ={1 \over 96}\bar\theta^{(\pm)} O_1\Pi^{(\pm)}\Gamma^{B_1B_2B_3}O_2
  \theta^{(\pm)}\bar\theta^{(\pm)}\Gamma_{B_1B_2B_3}\theta^{(\pm)}
\ee

\no where $\Pi^{(\pm)}={1 \over 2}(I\pm \Gamma_{(11)})$ are the Weyl projection
operators (see Appendix A for our conventions).  Then one obtains immediately
the vanishing of the triple contraction:

\be
\bar\theta^{(\pm)}\Gamma_{B_1B_2B_3}\theta^{(\pm)}\bar\theta^{(\pm)}
  \Gamma^{B_1B_2B_3}\theta^{(\pm)}=0
\ee

\no since, in 10 dimensions, $\Gamma_{B_1B_2B_3}\Gamma^{C_1C_2C_3}
\Gamma^{B_1B_2B_3}=-48\Gamma^{C_1C_2C_3}$.  Likewise, using the properties of
the Dirac algebra, it is relatively simple to show that the following double
contraction vanishes:

\be
\bar\theta^{(\pm)}\Gamma_{AB_1B_2}\theta^{(\pm)}\bar\theta^{(\pm)}
  \Gamma^{B_1B_2C}\theta^{(\pm)}=\bar\theta^{(\pm)}\Gamma_A\Gamma_{B_1B_2}
  \theta^{(\pm)}\bar\theta^{(\pm)}\Gamma^{B_1B_2}\Gamma^C\theta^{(\pm)}
  =0.
\ee

\no For the single trace we get a non-trivial result:

\be
\bar\theta^{(\pm)} \Gamma_{A_1A_2}\Gamma_B\theta^{(\pm)}
     \bar\theta^{(\pm)}\Gamma^{C_1C_2}\Gamma^B\theta^{(\pm)}
=2\bar\theta^{(\pm)}\Gamma_{B[A_1}\,\!^{[C_1}\theta^{(\pm)}
  \bar\theta^{(\pm)}\Gamma_{A_2]}\,\!^{C_2]B}\theta^{(\pm)}.
\ee

\no In particular, (2.4) implies the vanishing of the antisymmetric
combination:

\be
\bar\theta^{(\pm)}\Gamma_{[A_1A_2}\Gamma^B\theta^{(\pm)}\bar\theta^{(\pm)}
  \Gamma_{C_1C_2]}\Gamma_B\theta^{(\pm)}=0.
\ee

\no In fact, (2.4) implies the more powerful and useful result

\be
\bar\theta^{(\pm)}\Gamma_{[A_1A_2}\Gamma^B\theta^{(\pm)}\bar\theta^{(\pm)}
  \Gamma_{A_3]}\,\!^C\,\!_B\theta^{(\pm)}=0. \label{friend}
\ee

\no Therefore we conclude that $\bar\theta^{(\pm)}\Gamma_{A_1A_2}\Gamma_B
\theta^{(\pm)}\bar\theta^{(\pm)}\Gamma^{C_1C_2}\Gamma^B\theta^{(\pm)}$ is a
traceless tensor which contains no antisymmetric parts of more than 2 indices,
and must therefore correspond to the representation

\[\begin{picture}(30, 30)
\multiput(0, 0)(10, 0){2}{\framebox(10,10){ }}
\multiput(0, 10)(10, 0){2}{\framebox(10,10){ }}
\end{picture}
 \mbox{or} \;\; [2\;\;2].\]

\no Finally we are ready to tackle the uncontracted product, and we obtain:

\begin{eqnarray}
\lefteqn{{9 \over 8}\bar\theta^{(\pm)} \Gamma^{A_1A_2A_3}\theta^{(\pm)}
  \bar\theta^{(\pm)}\Gamma^{C_1C_2C_3}\theta^{(\pm)}= } \nonumber \\
 & & \mp{1 \over
32}\epsilon^{A_1A_2A_3D_1D_2D_3D_4D_5[C_1C_2}\bar\theta^{(\pm)}
  \Gamma^{C_3]}\,\!_{D_1D_2}\theta^{(\pm)}\bar\theta^{(\pm)}
  \Gamma_{D_3D_4D_5}\theta^{(\pm)} \nonumber \\
 & &-{9 \over 8}\bar\theta^{(\pm)}{\Gamma^{[A_1A_2}}^{\scriptstyle{[C_1}}
  \theta^{(\pm)}
  \bar\theta^{(\pm)}{\Gamma^{A_3]}}^{\scriptstyle{C_2C_3]}}\theta^{(\pm)}
\nonumber \\
 & & +{9 \over 4}{\eta^{[A_1}}^{\scriptstyle{[C_1}}
  \bar\theta^{(\pm)}\Gamma^{A_2A_3]}\,\!_D
  \theta^{(\pm)}\bar\theta^{(\pm)}{\Gamma^{D}}^{\scriptstyle{C_2C_3]}}
  \theta^{(\pm)}
\end{eqnarray}

\no where one has to make use of the Dirac algebra and in particular

\be
\Gamma^{A_1\ldots A_7}= {1 \over 3!} \epsilon^{A_1\ldots A_7B_1B_2B_3}
  \Gamma_{(11)}\Gamma_{B_1B_2B_3}.
\ee

\no Before we can make sense of Eq. (2.7), let us note that if we call:

\be
X^{(\pm)C;D_1\ldots D_5}=\bar\theta^{(\pm)}\Gamma^{C[D_1D_2}
  \theta^{(\pm)}\bar\theta^{(\pm)}\Gamma^{D_3D_4D_5]}\theta
\ee

\no we get
\begin{eqnarray}
\lefteqn{X^{(\pm)}_{[C_1;C_2C_3]}\,\!^{A_1A_2A_3}=} \nonumber \\
 & &{1 \over 10}
  \bigl(\bar\theta^{(\pm)}\Gamma^{A_1A_2A_3}\theta^{(\pm)}
\bar\theta^{(\pm)}\Gamma_{C_1C_2C_3}\theta^{(\pm)}
-3\bar\theta^{(\pm)}\Gamma^{[A_1A_2}\,\!_{[C_1}\theta^{(\pm)}
\bar\theta^{(\pm)}\Gamma^{A_3]}\,\!_{C_2C_3]}
\theta^{(\pm)}\bigr).
\end{eqnarray}

\no $X^{(\pm)}$ is clearly traceless by virtue of (2.5) and trivially satisfies

\be
X^{(\pm)[A;B_1\ldots B_5]}=0.
\ee

\no And, since $X^{(\pm)}$ has five totally antisymmetric indices, it is a good
candidate for the other irreducible piece of the $\theta^4$ sector.  This will
be confirmed shortly.  Then we can rewrite (2.7) as

\begin{eqnarray}
\lefteqn{\bar\theta^{(\pm)}\Gamma^{A_1A_2A_3}\theta^{(\pm)}
  \bar\theta^{(\pm)}\Gamma^{C_1C_2C_3}\theta^{(\pm)} = } \nonumber \\
  & &\mp{1 \over 48}\epsilon^{A_1A_2A_3D_1D_2D_3D_4D_5[C_1C_2}
  \bar\theta^{(\pm)C_3]}\,\!_{D_1D_2}\theta^{(\pm)}\bar\theta^{(\pm)}
  \Gamma_{D_3D_4D_5}\theta^{(\pm)}+\nonumber \\
  & &+{5 \over 2}X^{(\pm)[A_1;A_2A_3]C_1C_2C_3} \nonumber \\
 & &+{3 \over 2}\bar\theta^{(\pm)}
  \Gamma_B\,\!^{[A_2A_3}\theta^{(\pm)}
  \eta^{A_1][C_1}\bar\theta^{(\pm)}
  \Gamma^{C_2C_3]B}\theta^{(\pm)}.
\end{eqnarray}

\no This equation implies the (anti-) self-duality of $X^{(\pm)A;B_1\ldots
B_5}$:

\[X^{(\pm)A;B_1\ldots B_5}=\mp{1 \over 5!}
   \epsilon^{B_1\ldots B_5D_1
  \ldots D_5} X^{(\pm)A;}\,\!_{D_1\ldots D_5} \]
\be
X^{(\pm)A;}\,\!_{B_1\ldots B_5}=\pm {1 \over 5!}
  \epsilon_{B_1\ldots B_5D_1 \ldots D_5}X^{(\pm)A;D_1\ldots
D_5}
\ee

\no thus confirming that it is the missing irreducible piece from the
$\theta^4$ sector.

Therefore, the basic identity (2.12) gives the decomposition of the general
$\theta^4$ tensor in irreducible pieces.  It is the basic identity from which
all the higher order decompositions must necessarily follow by appropriate
iterative use of it.

In the remainder of the paper we are going to concentrate only on the positive
chirality case $\theta^{(+)}$.  To obtain the corresponding results for
$\theta^{(-)}$ one just has to remember that all the chirality and duality
properties are reversed.

\ve

\section{$\theta^6$ Decompositions}
\setcounter{equation}{0}

\b

In order to simplify notation let us call

\be
M^{ABC}=\bar\theta^{(+)} \Gamma^{ABC}\theta^{(+)}.
\ee

\no Also in the remainder of the paper we are going to use the following letter
convention: {\it uncontracted indices labeled by the same letter with
different subindex are understood to be antisymmetrized except if the letter
involved is $S$ or $X$ in which case they are understood to be symmetrized}.
For instance:

\[F^{CA_1A_2A_3}G^{A_4A_5D}\equiv F^{C[A_1A_2A_3}G^{A_4A_5]D} \]
\be
N^{CDS_1S_2}\,\!_{X_1}P^{S_3AB}\,\!_{X_2}\equiv N^{CD(S_1S_2}\,\!_{(X_1}
  P^{S_3)AB}\,\!_{X_2)}
\ee

\no where the square and round brackets are the by now standard notations
denoting normalized total antisymmetrization and symmetrization respectively.
This notation will dramatically reduce the need for brackets which would make
some formulae otherwise practically impossible to write.

Then, Eq. (2.12) becomes:
\begin{eqnarray}
\lefteqn{M^{A_1A_2A_3}M^{B_1B_2B_3}=} \nonumber \\
& &{5 \over 2}\biggl(M^{A_1[A_2A_3} M^{B_1B_2B_3]}
-{1 \over 5!}\epsilon^{A_1A_2A_3B_1B_2D_1\ldots D_5}
  M^{B_3}\,\!_{D_1D_2}M_{D_3D_4D_5}\biggr)\nonumber \\
 & &+{3 \over 2}\eta^{A_1B_1}
  M^{A_2A_3}\,\!_{D}M^{B_2B_3D}.
\end{eqnarray}

\no Eq. (3.3) is equivalent to the following two statements:

\be
M^{CA_1A_2}M^{A_3A_4A_5}=-{1 \over 5!} \epsilon^{A_1\ldots A_5B_1\ldots B_5}
  M^C\,\!_{B_1B_2}M_{B_3B_4B_5}
\ee

\be
M^{A_1A_2A_3}M^{B_1B_2B_3}=5M^{A_1[A_2A_3}M^{B_1B_2B_3]}+
  {3 \over 2}\eta^{A_1B_1}M^{A_2A_3}\,\!_DM^{B_2B_3D}.
\ee

\no Eqs. (3.3) or (3.5) clearly give the decomposition of $M^{A_1A_2A_3}
M^{B_1B_2B_3}$ into its irreducible parts, the anti-selfdual $[2 1 1 1 -1]$
piece:

\be
{\cal{M}}_4^{A;B_1\ldots B_5}=M^{AB_1B_2} M^{B_3B_4B_5}
\ee

\no and the $[2 2]$ piece:

\be
{\cal{M}}_4^{A_1A_2;B_1B_2}=M^{A_1A_2}\,\!_E M^{B_1B_2E}.
\ee

\no From their definitions and the results of this and the previous section, we
get the following properties:

\[{\cal{M}}_4^{[A;B_1\ldots B_5]}=0 \qquad
  {\cal{M}}_{4\,\,E}\,\!^{;EB_1\ldots B_4}=0 \]
\be
{\cal{M}}_4^{A;B_1\ldots B_5}=-{1 \over 5!}\epsilon^{B_1\ldots B_5
  D_1\ldots D_5} {\cal{M}}_4\,\!^{A;}\,\!_{D_1\ldots D_5}
\ee

\no and

\[{\cal{M}}_4^{A_1A_2;B_1B_2}={\cal{M}}_4^{B_1B_2;A_1A_2}
  \qquad {\cal{M}}_{4\,\,E}\,\!^{A;EB}=0 \]
\be
{\cal{M}}_4^{A[B;CD]}=0.
\ee

In order to decompose the next product
$M^{A_1A_2A_3}M^{B_1B_2B_3}M^{C_1C_2C_3}$ one can proceed to iterate (3.3) for
the different binary products.  After several iterations and a lot of algebra
it is possible to obtain the following decomposition:

\begin{eqnarray}
\lefteqn{M^{A_1A_2A_3} M^{B_1B_2B_3}M^{C_1C_2C_3}=} \nonumber \\
& &{\cal{S}}(A,B,C)\biggl\{18\eta^{B_1C_1}M^{[A_1A_2A_3}M^{B_2C_2]}\,\!_D
  M^{B_3C_3D} + {18 \over 5}\eta^{A_1C_1}\eta^{B_1C_2}M^{C_3}\,\!_{DE}
  M^{A_2A_3D}M^{B_2B_3E} \nonumber \\
& &-{9 \over 5}\eta^{B_1C_1}\eta^{B_2C_2}M^{C_3}\,\!_{DE}
  M^{B_3DA_1}M^{A_2A_3E}\biggr\} \nonumber \\
& &+{1 \over 20}\epsilon^{B_1B_2B_3A_1A_2C_1C_2D_1D_2D_3}M^{A_3E_1}\,\!_{D_1}
  M_{D_2D_3}\,\!^{E_2}M_{E_1E_2}\,\!^{C_3}
\end{eqnarray}

\no where ${\cal{S}}(A,B,C)$ is the normalized operator that fully symmetrizes
on the letters $A,B,C$.  The last term in (3.10) is automatically symmetric
upon interchange of these three letters, as can be easily proven by using the
fact that a complete antisymmetrization of 11 indices must necessarily vanish.

In deriving (3.10) one has to make use of many identities (see Appendix A)
which
are also consequences of (3.3), specially

\be
M^A\,\!_{DE} M^{BEF} M^C\,\!_F\,\!^D = 0
\label{nueva}
\ee

\no which follows almost immediately from (\ref{friend}) and (2.3).
Eq. (3.11) means that {\it all triple contractions of $M^3$ vanish}, as it
should be since there are no objects with 3 indices in the $\theta^6$ sector.

The amount of effort required to obtain (3.10) by iteration of (3.3) makes it
clear that an alternative way is needed if one hopes to decompose all the
higher order products.  Nevertheless it illustrates the fact that all the
necessary product decompositions are direct consequences of the Fierz identity
(2.12).

There is a much simpler way to obtain the decomposition (3.10), by
systematically removing traces (since the irreducible pieces are traceless) and
using the appropriate Young projectors on the traceless parts.  This is
possible because we already know beforehand what are the irreducible
representations involved (see Table 1).

Let us begin by removing all the traces from the object:

\begin{eqnarray}
\lefteqn{M^{A_1A_2A_3}M_D\,\!^{B_1B_2}M^{DC_1C_2}=\,\,Traceless\,\,
  (M^{A_1A_2A_3}M_D\,\!^{B_1B_2}M^{DC_1C_2})} \nonumber \\
& &+{2 \over 5}\bigl(2\eta^{A_1B_1}M^{EA_2A_3}M_{DE}\,\!^{B_2} M^{DC_1C_2}
+2\eta^{A_1C_1}M^{EA_2A_3}M^{DB_1B_2}M_{DE}\,\!^{C_2} \nonumber \\
& &+\eta^{B_1C_1}M^{EA_2A_3}M_{DE}\,\!^{B_2}M^{DC_2A_1}\bigr)
\end{eqnarray}

\no Next we decompose $Traceless\,(M^{A_1A_2A_3}M_D\,\!^{B_1B_2}M^{DC_1C_2})$
using the Young projectors corresponding to the representation
\setlength{\unitlength}{.5pt}
$\begin{picture}(20, 30)
\multiput(0, 20)(0, -10){5}{\framebox(10,10){ }}
\multiput(10, 20)(0, -10){2}{\framebox(10,10){ }}
\end{picture}$
(see Table 1) whose construction is detailed in Appendix C:

\begin{eqnarray}
\lefteqn{Traceless\, \bigl(M^{A_1A_2A_3}M_D\,\!^{B_1B_2}M^{DC_1C_2}\bigr)=}
\nonumber \\
& &= Y\left(
\setlength{\unitlength}{.5pt}
\begin{picture}(20, 40)
\multiput(0, 20)(0, -10){5}{\framebox(10,10){ }}
\multiput(10, 20)(0, -10){2}{\framebox(10,10){ }}
\end{picture}
\right)M^{A_1A_2A_3}M^{B_1B_2D}M^{C_1C_2}\,\!_D \nonumber \\
&  &={2 \over 3}\bigl(M^{[A_1A_2A_3}M^{B_1B_2]}\,\!_DM^{C_1C_2D}
+ M^{[A_1A_2A_3}M^{C_1C_2]}\,\!_D M^{B_1B_2D} \nonumber \\
&  &+2M^{[A_1A_2A_3}M^{B_1C_1]}\,\!_D M^{B_2C_2D}\bigr).
\end{eqnarray}

Now we do the same for the uncontracted product $M^{A_1A_2A_3}M^{B_1B_2B_3}
M^{C_1C_2C_3}$, first remove the traces:

\begin{eqnarray}
\lefteqn{M^{A_1A_2A_3}M^{B_1B_2B_3}M^{C_1C_2C_3}=\,Traceless\,
  \bigl(M^{A_1A_2A_3}M^{B_1B_2B_3}M^{C_1C_2C_3}\bigr)} \nonumber \\
& &+{9 \over 5}{\cal{S}}(A,B,C)\biggr\{\eta^{A_1B_1}
  \biggl[{3 \over 2}M_D\,\!^{A_2A_3}M^{DB_2B_3}M^{C_1C_2C_3}
-M_D\,\!^{A_2A_3}M^{C_1B_2B_3}M^{DC_2C_3} \nonumber \\
& &+2M_D\,\!^{A_2A_3}M^{C_1C_2B_2} M^{DB_3C_3}
  +2M_D\,\!^{A_2B_2}M^{DC_1C_2}M^{A_3B_3C_3}\biggr]\biggr\}.
\end{eqnarray}

\no Using some of the identities in Appendix A and the decomposition
(3.12)-(3.13) we get

\begin{eqnarray}
\lefteqn{M^{A_1A_2A_3}M^{B_1B_2B_3}M^{C_1C_2C_3}=\,Traceless\,
  \bigl(M^{A_1A_2A_3}M^{B_1B_2B_3}M^{C_1C_2C_3}\bigr)} \nonumber \\
  & &+9{\cal{S}}(A,B,C)\biggl\{2\eta^{A_1B_1}M^{[C_1C_2C_3}M^{A_2B_2]}\,\!_D
  M^{A_3B_3D} +{2 \over 5}\eta^{A_1B_1}\eta^{C_1B_2}M^{B_3}\,\!_{DE}
  M^{A_2A_3D}M^{C_2C_3E} \nonumber \\
  & &-{1 \over 5}\eta^{A_1B_1}\eta^{A_2B_2}M^{B_3}\,\!_{DE}M^{C_1C_2D}
  M^{B_3C_3E}\biggr\}.
\end{eqnarray}

\no To obtain the traceless part in (3.15), we apply the Young projector
corresponding to the representation
\setlength{\unitlength}{.5pt}
$\begin{picture}(30, 50)
\multiput(0, 30)(0, -10){7}{\framebox(10,10){ }}
\multiput(10, 30)(10, 0){2}{\framebox(10,10){ }}
\end{picture}$
($\equiv
\begin{picture}(30, 30)
\multiput(0, 10)(0, -10){3}{\framebox(10,10){ }}
\multiput(10, 10)(10, 0){2}{\framebox(10,10){ }}
\end{picture}$ for $SO(10)$)

\begin{eqnarray}
\lefteqn{Traceless\,\bigl(M^{A_1A_2A_3}M^{B_1B_2B_3}M^{C_1C_2C_3}\bigr)
= \mbox{       }}\nonumber \\
& &=Y\left(
\setlength{\unitlength}{.5pt}
\begin{picture}(30, 50)
\multiput(0, 30)(0, -10){7}{\framebox(10,10){ }}
\multiput(10, 30)(10, 0){2}{\framebox(10,10){ }}
\end{picture}
\right)M^{A_1A_2A_3}M^{B_1B_2B_3}M^{C_1C_2C_3}\nonumber \\
& &=21M^{B_1[B_2B_3}M^{A_1A_2A_3}M^{C_2C_3]C_1}\nonumber \\
& &=-{1 \over 20} \epsilon^{A_1A_2A_3B_1B_2C_1C_2E_1E_2E_3}
  M^{FDB_3}M_{DE_1E_2}M_{E_3F}\,\!^{C_3}
\end{eqnarray}

\no
where the last equality follows from the anti-selfduality
of $M^{A[B_1B_2}M^{B_3B_4B_5]}$ by rotating indices, and explicitly displays
the aforementioned equivalence of $SO(10)$ representations.

Eq. (3.16) together with (3.15) reproduces for us the decomposition (3.10).  We
will delay the study of the irreducible pieces of the $\theta^6$ sector until
the next section.
\ve

\section{Irreducible Bosonic Structures}
\setcounter{equation}{0}

\b

The difficulty in proceeding along the lines of the previous section is that
one needs to know beforehand what are the irreducible pieces of the higher
$\theta$ powers in order to decompose the products into irreducible pieces.
That is why we are now going to proceed {\it backwards}, starting from
the scalar corresponding to $\theta^{16}$ and come down from there.

To construct the above scalar we first notice that it is easy to identify the
totally symmetric tensor of $\theta^8$ sector corresponding to the
representation [4]:

\be
{\cal{M}}_8^{ABCD}=M^A\,\!_E\,\!^FM^B\,\!_F\,\!^GM^C\,\!_G\,\!^H
  M^D\,\!_H\,\!^E.
\ee

\no It is obviously traceless (see (\ref{nueva})and cyclically symmetric:

\be
{\cal{M}}_8^{ABCD} = {\cal{M}}_8^{DABC}
\ee

\no and the antisymmetrization of any two neighboring indices vanishes

\begin{eqnarray}
{\cal{M}}_8^{[AB]CD} & = & M^{[A}\,\!_{EF}M^{B]FG}M^C\,\!_G\,\!^H
  M^D\,\!_H\,\!^E \nonumber \\
&  &=-{1 \over 2}M^{BA}\,\!_FM_E\,\!^{FG}M^C\,\!_{GH}M^{DHE} \nonumber \\
&  &={1 \over 4} M^{BA}\,\!_FM_{HE}\,\!^GM^C\,\!_G\,\!^FM^{DHE}=0
\end{eqnarray}

\no where we have twice made use of (2.6) and then (2.3).  Thus

\be
{\cal{M}}_8^{ABCD}={\cal{M}}_8^{BACD}.
\ee

\no Properties (4.2) and (4.4) imply that ${\cal{M}}_8^{ABCD}$ is completely
symmetric in all four indices.

\vskip.5cm
\no${\underline{\theta^{16}}}$.

The scalar we are looking for is the square of (4.1)

\begin{eqnarray}
{\cal{M}}_{16}&=&{\cal{M}}_8^{S_1S_2S_3S_4}{\cal{M}}_{8\,\,  S_1S_2S_3S_4}
\nonumber \\
  &=&M^{S_1}\,\!_{E_1E_2}M^{S_2E_2F_1}M^{S_3}\,\!_{F_1F_2}M^{S_4F_2E_1}
M_{S_1}\,\!^{G_1G_2}M_{S_2G_2H_1}M_{S_3}\,\!^{H_1H_2}M_{S_4H_2G_1}
\end{eqnarray}

\no where all the $M$-factors are equivalent.

\vskip.5cm
\no${\underline{\theta^{14}}}$.

Since all the factors in (4.5) are equivalent, there is only one possible
expression to be obtained by removing any one of them and that must be our
irreducible piece:

\begin{eqnarray}
{\cal{M}}^{ABC}&=&M^{S_1}\,\!_{E_1E_2}M^{S_2E_2F_1}
M^{S_3}\,\!_{F_1F_2}M^{AF_2E_1}M_{S_1}\,\!^{BG}M_{S_2GH}M_{S_3}\,\!^{HC}
\nonumber \\
  &=&{\cal{M}}_8^{S_1S_2S_3A}M_{S_1}\,\!^{BG}M_{S_2GH}
   M_{S_3}\,\!^{HC}
\end{eqnarray}

\no which is obviously antisymmetric in $B,C$:

\be
{\cal{M}}^{ABC}=-{\cal{M}}^{ACB}
\ee

\no but must be totally antisymmetric because it must belong to
\setlength{\unitlength}{.75pt}
$\begin{picture}(10, 30)
\multiput(0, 10)(0, -10){3}{\framebox(10,10){ }}
\end{picture} \equiv [1 \;1 \; 1]$  .  In order to
prove this, we first put it in a more appealing form using the symmetry of the
$\begin{picture}(40, 0)
\multiput(0, 0)(10, 0){4}{\framebox(10,10){ }}
\end{picture}$ part as well as (\ref{friend}):

\be
{\cal{M}}^{ABC}=M^{AS_1D_1}M^{S_2} \,\!_{D_1D_2}M^{ED_2F_1}M^B \,\!_{F_1F_2}
  M_{S_1}\,\!^{F_2G_1}M_{S_2G_1G_2}M^C\,\!_E\,\!^{G_2}.
\ee

\no Then, reordering factors and using (\ref{friend}) once more we obtain

\begin{eqnarray}
{\cal{M}}^{ABC}&=&M^{BS_1D_1}M^{S_2}\,\!_{D_1D_2}M^{ED_2F_1}
  M^C \,\!_{F_1F_2}M_{S_1}\,\!^{F_2G_1}M_{S_2G_1G_2}M^A\,\!_E\,\!^{G_2}
\nonumber \\
  &=&{\cal{M}}^{BCA}.
\end{eqnarray}

Properties (4.7) and (4.9) imply that ${\cal{M}}^{ABC}$ is completely
antisymmetric in all 3 indices. From (4.6) and (4.5) we note that

\be
{\cal{M}}^{A_1A_2A_3}M_{A_1A_2A_3}=-{\cal{M}}_{16}
\ee

\no and therefore we have the product decomposition

\be
{\cal{M}}^{A_1A_2A_3} M^{B_1B_2B_3}=-{1 \over 120}
  \eta^{A_1B_1}\eta^{A_2B_2}\eta^{A_3B_3}{\cal{M}}_{16}.
\ee
\vskip.5cm

\no${\underline{\theta^{12}}}$.

Not all the factors in (4.6) are equivalent, so now we get two possible
structures by removing one factor from ${\cal{M}}^{ABC}$.  One is:

\be
\hat{\cal{M}}_{12}\,\!^{AB,CD}=M^{AS_1}\,\!_{D_1}\,\!M^{S_2D_1D_2}M^C\,\!
  _{D_2E}M^{DEF_1}M_{S_1F_1F_2}M_{S_2}\,\!^{F_2B}
\ee

\no which is clearly traceless and, by virtue of (\ref{friend}), (2.3), has the
symmetry properties:

\be
\hat{\cal{M}}_{12}\,\!^{AB,CD}=\hat{\cal{M}}_{12}\,\!^{BA,CD}=
  \hat{\cal{M}}_{12}\,\!^{AB,DC}=\hat{\cal{M}}_{12}\,\!^{BA,DC}.
\ee

\no By using (\ref{friend}) in a different way we can also derive

\be
\hat{\cal{M}}_{12}\,\!^{AB,CD}+\hat{\cal{M}}_{12}\,\!^{AC,DB}
  +\hat{\cal{M}}_{12}^{CB,AD}=0
\ee

\be
\hat{\cal{M}}_{12}\,\!^{AB,CD}+\hat{\cal{M}}_{12}\,\!^{DB,AC}
  +\hat{\cal{M}}_{12}^{AD,CB}=0.
\ee

\no Combining (4.14) with (4.13) we get

\be
\hat{\cal{M}}_{12}\,\!^{A[B,C]D}+\hat{\cal{M}}_{12}\,\!^{D[B,C]A}
  =0
\ee

\no while combining (4.14) and (4.15),

\be
\hat{\cal{M}}_{12}\,\!^{AB,CD}=\hat{\cal{M}}_{12}\,\!^{CD,AB}.
\ee

\no Once we have obtained (4.17) we see that (4.14) and (4.15) simply mean:

\be
\hat{\cal{M}}_{12}\,\!^{A(B,CD)}=0.
\ee

Eq. (4.16) tells us that antisymmetrizing on two indices on opposite sides of
the comma automatically makes the other pair also antisymmetric.  Thus we
recognize the object that displays the symmetry of the Young pattern
\setlength{\unitlength}{.5pt}
$\begin{picture}(30, 30)
\multiput(0, 0)(10, 0){2}{\framebox(10,10){ }}
\multiput(0, 10)(10, 0){2}{\framebox(10,10){ }}
\end{picture}$:

\be
{\cal{M}}_{12}\,\!^{A_1A_2;B_1B_2}=\hat{\cal{M}}_{12}\,\!^{A_1B_1,B_2A_2}
= \hat{\cal{M}}_{12}\,\!^{A_1B_1,A_2B_2}.
\ee

\no However it is interesting to note for reference, the more interesting
properties of the $\hat{\cal{M}}_{12}$ tensor.  From the definition (4.19) it
is clear that $\hat{\cal{M}}_{12}^{A_1A_2,B_1B_2}$ is traceless and that it
satisfies:

\be
{\cal{M}}_{12}\,\!^{A[B;CD]}=0.
\ee

\no Thus it has the same properties as the tensor
${\cal{M}}_{12}\,\!^{A_1A_2;B_1B_2}$ except for nilpotency.

Even though $\hat{\cal{M}}_{12}$ and ${\cal{M}}_{12}$ have apparently different
symmetry properties they both have the same number of degrees of freedom, 770,
i.e. the dimension of the irrep. $[2 2]$ of $SO(10)$, and they both can be
expressed in terms of the other.  The inverse of (4.19) is

\be
\hat{\cal{M}}_{12}\,\!^{AB,CD}={2 \over 3}
\bigl({\cal{M}}_{12}\,\!^{AD;BC}
+{\cal{M}}_{12}\,\!^{BD;AC}\bigr)
\ee

\no as can be easily seen by using (4.18).

{}From (4.8) and (4.12) we see that

\be
\hat{\cal{M}}_{12}^{AE,BF}M^C\,\!_{FE}=
  {\cal{M}}_{12}\,\!^{AB;EF}M^C\,\!_{FE}={\cal{M}}^{ABC}
\ee

\no and then we have for the decomposition of the single contraction:

\be
\hat{\cal{M}}_{12}\,\!^{S_1S_2,XE}M_{A_1A_2E}={1 \over 7}
  \biggl(3\delta^{S_1}_{A_1}{\cal{M}}^{S_2X}\,\!_{A_2}-
  {1 \over 3}\eta^{XS_1}{\cal{M}}^{S_2}\,\!_{A_1A_2}+{1 \over 3}
  \eta^{S_1S_2}{\cal{M}}^X\,\!_{A_2A_1}\biggr).
\ee

\no Eq. (4.23) is easily obtained since it must have that general form and the
coefficients are given by the traces of the left-hand side, either zero or
(4.22).  For the other object we have

\be
{\cal{M}}_{12}\,\!^{B_1B_2;CE}M_E\,\!^{A_1A_2}={1 \over 14}
  \bigl(3\eta^{A_1C}{\cal{M}}^{A_2B_1B_2}-3\eta^{A_1B_1}{\cal{M}}^{A_2B_2C}
  +\eta^{CB_1}{\cal{M}}^{B_2A_1A_2}\bigr).
\ee

\no Using (4.23) and following the same procedure one derives for the full
product

\begin{eqnarray}
\hat{\cal{M}}_{12}\,\!^{S_1S_2,X_1X_2}M^{A_1A_2A_3} &=&
\frac{3}{11 \times 7}\biggl[8\eta^{S_1A_1}\eta^{X_1A_2}{\cal{M}}^{S_2X_2A_3}
\nonumber \\
& &-\eta^{S_1X_1}\eta^{S_2A_1}{\cal{M}}^{X_2A_2A_3}
-\eta^{S_1X_1}\eta^{X_2A_1}{\cal{M}}^{S_2A_2A_3}
\nonumber \\
& &+\eta^{X_1X_2}\eta^{S_1A_1}{\cal{M}}^{S_2A_2A_3}
+\eta^{S_1S_2}\eta^{X_1A_1}{\cal{M}}^{X_2A_2A_3}
\nonumber \\
& &-\frac{1}{9}\left(\eta^{X_1X_2}\eta^{S_1S_2}
- \eta^{X_1S_1}\eta^{X_2S_2} \right){\cal{M}}^{A_1A_2A_3} \biggr]
\end{eqnarray}

\no and

\begin{eqnarray}
{\cal{M}}_{12}\,\!^{B_1B_2;C_1C_2}M^{A_1A_2A_3}& =&
-{6 \over 11 \times 7}\biggl(\eta^{A_1B_1}\eta^{A_2B_2}{\cal{M}}^{A_2C_1C_2}
  +2\eta^{A_1B_1}\eta^{A_2C_2}{\cal{M}}^{A_3B_2C_2} \nonumber \\
& +&\eta^{A_1C_1}\eta^{A_2C_2}{\cal{M}}^{A_3B_1B_2}-{3 \over 4}
  \eta^{B_1C_1}\eta^{C_2A_1}{\cal{M}}^{A_2A_3B_2}\nonumber \\
& -&{3 \over 4}\eta^{B_1C_1}\eta^{B_2A_1}{\cal{M}}^{A_2A_3C_2}
  +{1 \over 12}\eta^{B_1C_1}\eta^{B_2C_3}{\cal{M}}^{A_1A_2A_3}\biggl).
\end{eqnarray}

\no If we remove a different factor from ${\cal{M}}^{ABC}$ we extract the new
structure

\be
\hat{\cal{M}}_{12}\,\!^{XABY;E_1E_2}=M^X\,\!_{D_1}\,\!^{D_2}
  M^A\,\!_{D_2}\,\!^{D_3}M^F\,\!_{D_3}\,\!^{D_4}M^B\,\!_{D_4}
  \,\!^{D_5}M^Y\,\!_{D_5}\,\!^{D_1}M_F\,\!^{E_1E_2}.
\ee

\no It has the obvious property

\be
\hat{\cal{M}}_{12}^{XABY;C_1C_2}=-\hat{\cal{M}}_{12}^
  {YBAX;C_1C_2}
\ee

\no and by applying (\ref{friend}) it is also easy to prove

\be
\hat{\cal{M}}_{12}^{X[ABY;C_1C_2]}=
\hat{\cal{M}}_{12}^{[A^{\scriptstyle X}BY;C_1C_2]}
\ee

\no which in turn implies:

\be
\hat{\cal{M}}_{12}^{[XABY;C_1C_2]}=0.
\ee

However, this object is not irreducible because it is not completely traceless,
but rather has two non-vanishing traces:

\[\hat{\cal{M}}_{12\,\,E}\,\!^{ABY;EC}=-\hat{\cal{M}}_{12}\,\!
  ^{AC,BY} \]
\be
  \hat{\cal{M}}_{12}\,\!^{XAB}\,\!_E\,\!^{;EC}=\hat{\cal{M}}_{12}\,\!
  ^{BC,AX}.
\ee

\no In order to decompose it one removes the traces and applies the appropriate
Young projector:

\begin{eqnarray}
\lefteqn{\hat{\cal{M}}_{12}\,\!^{XABY;C_1C_2}=\,Traceless\,
  \bigl(\hat{\cal{M}}_{12}\,\!^{XABY;C_1C_2}\bigr)} \nonumber \\
& &+{1 \over 9\times 21}\biggl\{-46\bigl(\eta^{XC_1}\hat{\cal{M}}_{12}
  ^{AC_2,BY}-\eta^{YC_1}\hat{\cal{M}}_{12}\,\!^{BC_3,AX}\bigr) \nonumber \\
& &-3\bigl(\eta^{XB}\hat{\cal{M}}_{12}\,\!^{AC_1,YC_2}-\eta^{YA}
  \hat{\cal{M}}_{12}\,\!^{BC_1,XC_2}\bigr) \nonumber \\
& &-5\bigl(\eta^{XC_1}\hat{\cal{M}}_{12}\,\!^{BC_2,AY}-\eta^{YC_1}
  \hat{\cal{M}}_{12}\,\!^{AC_1,BX}+\eta^{AC_1}\hat{\cal{M}}_{12}\,\!^{BC_2,XY}
\nonumber \\
& &-\eta^{BC_1}\hat{\cal{M}}_{12}\,\!^{AC_2,XY}
  -\eta^{AC_1}\hat{\cal{M}}_{12}\,\!^{SC_2,BY}+\eta^{BC_1}
  \hat{\cal{M}}_{12}\,\!^{YC_1,AX}\bigr) \nonumber \\
& &+2\bigl(\eta^{XA}\hat{\cal{M}}_{12}^{BC_1,YC_2}-\eta^{YB}
  \hat{\cal{M}}_{12}\,\!^{AC_1,XC_2} \nonumber \\
& &+4\eta^{XY}\hat{\cal{M}}_{12}\,\!
  ^{AC_1,BC_2}-\eta^{AB}\hat{\cal{M}}_{12}\,\!^
  {XC_1,YX_2}\bigr)\biggr\},
\end{eqnarray}

\ve

\begin{eqnarray}
Traceless\,(\hat{\cal{M}}_{12}\,\!^{XABY;C_1C_2})
  &=&Y\left(
\setlength{\unitlength}{.5pt}
\begin{picture}(20, 40)
\multiput(0, 20)(0, -10){5}{\framebox(10,10){ }}
\put(10,20){\framebox(10,10){ }}
\end{picture}
\right)\hat{\cal{M}}_{12}\,\!^{XABY;C_1C_2} \nonumber \\
 &= &{5 \over 6}\bigl(\hat{\cal{M}}_{12}\,\!^{X[ABY;C_2C_3]}
  +\hat{\cal{M}}_{12}\,\!^{A[BYX;C_1C_2]} \nonumber \\
 & &+\hat{\cal{M}}_{12}\,\!^{B[YXA;C_1C_2]}
  +\hat{\cal{M}}_{12}\,\!^{Y[XAB;C_1C_2]}\bigr)
\end{eqnarray}

{}From (4.33) it is apparent that the second irreducible structure is

\begin{eqnarray}
{\cal{M}}_{12}^{B;A_1\ldots A_5}&=& \hat{\cal{M}}_{12}^{BA_1A_2A_3;A_4A_5}
\nonumber \\
  &=&M^B\,\!_{D_1}\,\!^{D_2}M^F\,\!_{D_2}\,\!^{D_3}M^{A_1}\,\!_{D_3}\,\!^{D_4}
  M^{A_2}\,\!_{D_4}\,\!^{D_5}M^{A_3}\,\!_{D_5}\,\!^{D_1}M_F \,\!^{A_4A_5},
\end{eqnarray}

\no whose tracelessness is confirmed by (\ref{nueva}).
Eq. (4.30) implies the property

\be
{\cal{M}}_{12}^{[B;A_1\ldots A_5]}=0
\ee

\no and in Appendix A we prove the duality property

\be
{\cal{M}}_{12}^{B;A_1\ldots A_5}= {1 \over 5!}
  \epsilon^{A_1\ldots A_5 E_1\ldots E_5}{\cal{M}}^{B;}_{12 \,\, E_1\ldots E_5}
  \ee

\no which is opposite to the one satisfied by ${\cal{M}}_4\,\!^{B;A_1\ldots
A_5}$.  The definitions (4.34), (4.6) give the result for the triple
contractions

\[{\cal{M}}_{12}^{B;A_1A_2E_1E_2E_3}M_{E_1E_2E_3}=-{1 \over 5}
  {\cal{M}}^{BA_1A_2} \]
\be
  {\cal{M}}_{12}^{E_1;E_2E_3A_1A_2A_3}M_{E_1E_2E_3}=-{1 \over 5}
  {\cal{M}}^{A_1A_2A_3}.
\ee

\no and by simple detracing,

\begin{eqnarray}
{\cal{M}}_{12}\,\!^{B;A_1A_2A_3E_1E_2}M_{CE_1E_2}&=&-{1 \over 70}
  \bigl(\delta_C^B{\cal{M}}^{A_1A_2A_3}-\eta^{BA_1}{\cal{M}}_C^{A_2A_3}
  +5\delta_C^{A_1}{\cal{M}}^{BA_2A_3}\bigr) \nonumber \\
  {\cal{M}}_{12}^{E_1;E_2A_1\ldots A_1}M_{CE_1E_2}&=&-{4 \over 35}
  \delta_C^{A_1}{\cal{M}}^{A_2A_3A_4}.
\end{eqnarray}

\no From (4.38) and Young-projecting

\begin{eqnarray}
\lefteqn{{\cal{M}}_{12}\,\!^{B;A_1\ldots A_4E}M_{C_1C_2E} =\,Traceless\,
  \bigl({\cal{M}}_{12}\,\!^{B;A_1\ldots A_4E}M_{C_1C_2E}\bigr) } \nonumber \\
  & &+{2 \over 3\times 35}\bigl(\delta^B_{C_1}{\cal{M}}^{A_2A_3A_4}
  -2\delta^{A_1}_{C_1}\delta^{A_2}_{C_2}{\cal{M}}^{BA_3A_4}
-\eta^{BA_1}\delta^{A_2}_{C_1}{\cal{M}}_{C_2}\,\!^{A_3A_4}  \bigr)
\end{eqnarray}

\begin{eqnarray}
Traceless\,\bigl({\cal{M}}_{12}\,\!^{B;A_1\ldots A_4E}
  M^{C_1C_2}\,\!_E\bigr)&=&Y\left(
\setlength{\unitlength}{.5pt}
\begin{picture}(10, 50)
\multiput(0,30)(0,-10){7}{\framebox(10,10){ }}
\end{picture}
\right){\cal{M}}_{12}\,\!^{B;A_1\ldots A_4E}
  M^{C_1C_2}\,\!_E \nonumber \\
 & =&{\cal{M}}_{12}\,\!^{[B;A_1\ldots A_4}\,\!_EM^{C_1C_2]E}
  ={1 \over 5}{\cal{M}}_{12}\,\!^{E;[BA_1\ldots A_4}M^{C_1C_2]}\nonumber \\
 & =&-{4 \over 5}{1 \over 7!}\epsilon^{BA_1\ldots A_4C_1C_2E_1E_2E_3}
  {\cal{M}}_{E_1E_2E_3}.
\end{eqnarray}

\no Eqs. (4.39), (4.40) and (4.35) then give

\be
{\cal{M}}_{12}^{E;A_1\ldots A_5}M^{C_1C_2}\,\!_E=-{2 \over 21}
  \biggl({1 \over 5!}\epsilon^{A_1\ldots A_5C_1C_2E_1E_2E_3}
  {\cal{M}}_{E_1E_2E_3}+\eta^{A_1C_1}\eta^{A_2C_2}
  {\cal{M}}^{A_3A_4A_5}\biggr).
\ee

\no Finally for the full product

\begin{eqnarray}
{\cal{M}}_{12}\,\!^{B;A_1\ldots A_5}M^{C_1C_2C_3} &=&
  -{2 \over 7!}\bigl\{\eta^{BC_1}\epsilon^{A_1\ldots A_5C_2C_3E_1E_2E_3}
  -\eta^{BA_1}\epsilon^{A_2\ldots A_5C_1C_2C_3E_1E_2E_3}
\nonumber \\
  &\mbox{ }&+\eta^{C_1A_1}
  \epsilon^{BA_2\ldots A_5C_2C_3E_1E_2E_3}\bigr\}{\cal{M}}_{E_1E_2E_3}
\nonumber \\
  &-&{1 \over 35}\biggl[\eta^{BC_1}\eta^{A_1C_2}\eta^{A_2C_3}
  {\cal{M}}^{A_3A_4A_5}+\eta^{A_1C_1}\eta^{A_2C_2}\eta^{A_3C_3}
  {\cal{M}}^{BA_4A_5} \nonumber \\
  &-&\eta^{BA_1}\eta^{A_2C_1}\eta^{A_3C_2}
  {\cal{M}}^{A_4A_5C_3}\biggr].
\end{eqnarray}

\no$\underline{\theta^{10}}$.

The first structure we encounter by removing a factor from
$\hat{\cal{M}}_{12}\,\!^{AB,CD}$ is

\be
\hat{\cal{M}}_{10}\,\!^{S_1S_2S_3;A_1A_2}={\cal{M}}_8^{S_1S_2S_3E}
  M^{A_1A_2}\,\!_E
\ee

\no whose symmetry properties are manifest.  Its tracelessness follows from
these symmetries and from the tracelessness of
${\cal{M}}_8\,\!^{S_1S_2S_3S_4}$.  The object in (4.43) also satisfies

\be
\hat{\cal{M}}_{10}\,\!^{(S_1S_2S_3;A)B}=0,
\ee

\no
and its product decompositions can be derived as before and we just list them:

\[\hat{\cal{M}}_{10}\,\!^{EDA;BF}M^C\,\!_{EF} =
  -\hat{\cal{M}}_{12}\,\!^{AD,BC} \]

\[  \hat{\cal{M}}_{10}\,\!^{ABD;EF}M^C\,\!_{EF}=\hat{\cal{M}}_{10}\,\!^{EFA;BD}
  M^C\,\!_{EF}=0 \]

\[\hat{\cal{M}}_{10}\,\!^{S_1S_2S_3;AE}M^{B_1B_2}\,\!_E =
  {4 \over 7}\eta^{S_1B_1}\hat{\cal{M}}_{12}\,\!^{S_2S_3,AB_2}
  -{2 \over 21}\eta^{S_1S_2}\hat{\cal{M}}_{12}\,\!^{S_3B_1,AB_2}\]

\begin{eqnarray}
\hat{\cal{M}}_{10}\,\!^{S_1S_2E;A_1A_2}M^{B_1B_2}\,\!_E &=&
  -{10 \over 3}{\cal{M}}_{12}\,\!^{S_1;S_2A_1A_2B_1B_2}
 \nonumber \\
&+&{2 \over 63}\biggl[{10 \over 3}\bigl(\eta^{S_1B_1}
\hat{\cal{M}}_{12}\,\!^{A_1B_2,A_2S_2}
  +\eta^{S_1A_1}\hat{\cal{M}}_{12}\,\!^{A_2B_1,B_2S_2}\bigr)\nonumber \\
&+& 17\eta^{A_1B_1}\hat{\cal{M}}_{12}\,\!^{S_1S_2,A_2B_2}
+{2 \over 3}\eta^{S_1S_2}\hat{\cal{M}}_{12}\,\!^{A_1B_1,A_2B_2}\biggr]
\nonumber \\
\nonumber \\
\hat{\cal{M}}_{10}\,\!^{S_1S_2S_3;A_1A_2}M^{B_1B_2B_3}&=&
  -{15 \over 11}\biggl\{{18 \over 7}\eta^{S_1B_1}{\cal{M}}_{12}\,\!^
  {S_2;S_3A_1A_2B_2B_3}
 -\eta^{S_1S_2}{\cal{M}}_{12}\,\!^{B_1;S_3A_1A_2B_2B_3}\biggr\} \nonumber \\
  &+&{4 \over 7}\bigl(\eta^{S_1A_1}{\cal{M}}_{12}\,\!^{S_2;S_3A_2B_1B_2B_3}
  -\eta^{S_1S_2}{\cal{M}}_{12}\,\!^{A_1;S_3A_2B_1B_2B_3}\bigr) \nonumber \\
  &+&{2 \over 35}\biggl[2\eta^{S_1B_1}\eta^{S_2A_1}\hat{\cal{M}}_{12}\,\!^
  {A_2B_2,B_3S_3}+9\eta^{S_1B_1}\eta^{B_2A_1}\hat{\cal{M}}_{12}\,\!^
  {S_2S_3,A_2B_3}\biggr] \nonumber \\
  &-&{1 \over 21}\biggl[2\eta^{S_1S_2}\eta^{A_1B_1}\hat{\cal{M}}_{12}\,\!^
  {A_2B_2,B_3S_3}-{1 \over 5}\eta^{S_1S_2}\eta^{S_3B_1}
  \hat{\cal{M}}_{12}\,\!^{A_1B_2,A_2B_3}\biggr].
\nonumber \\
\end{eqnarray}

However, the symmetry properties of the tensor
$\hat{\cal{M}}_{10}\,\!^{S_1S_2S_3;A_1A_2}$ are not the ones of the Young
pattern
\setlength{\unitlength}{.5pt}
$\begin{picture}(30, 30)
\multiput(0, 10)(0, -10){3}{\framebox(10,10){ }}
\multiput(10, 10)(10, 0){2}{\framebox(10,10){ }}
\end{picture}$
as it is conventionally understood, but it is easy to
construct a new tensor which corresponds to
\setlength{\unitlength}{.5pt}
$\begin{picture}(30, 30)
\multiput(0, 10)(0, -10){3}{\framebox(10,10){ }}
\multiput(10, 10)(10, 0){2}{\framebox(10,10){ }}
\end{picture}$ :

\be
{\cal{M}}_{10}\,\!^{S_1S_2;A_1A_2A_3}=\hat{\cal{M}}_{10}\,\!^
  {S_1S_2[A_1;A_2A_3]}.
\ee

\no But, just like we had in the $\theta^{12}$ case, both of these objects are
equivalent, both are irreducible and carry the same number of degrees of
freedom (4312) and they can be expressed in terms of each other.  The inverse
of (4.46) is:

\be
\hat{\cal{M}}_{10}\,\!^{S_1S_2B;A_1A_2}={3 \over 5}
  \bigl({\cal{M}}_{10}\,\!^{S_1S_2;BA_1A_2}
  +2{\cal{M}}_{10}\,\!^{S_1B;S_2A_1A_2}\bigr).
\ee

\no From the definition (4.46) we get the property

\be
{\cal{M}}_{10}\,\!^{S[B;A_1A_2A_3]}=0.
\ee

\no The new products are immediately obtained from (4.45):

\[{\cal{M}}_{10}\,\!^{SE_1;A_1A_2E_2}M^C\,\!_{E_1E_2}=
  -{2 \over 3} {\cal{M}}_{12}\,\!^{A_1A_2;SC}\]

\[{\cal{M}}_{10}\,\!^{S_1S_2;AE_1E_2}M^B\,\!_{E_1E_2}
  ={2 \over 3}\hat{\cal{M}}_{12}\,\!^{S_1S_2,AB}={8 \over 9}
  {\cal{M}}_{12}^{S_1B;S_2A}\]

\[{\cal{M}}_{10}\,\!^{E_1E_2;A_1A_2A_3}M^B\,\!_{E_1E_2}=0\]

\begin{eqnarray*}
{\cal{M}}_{10}\,\!^{SE;A_1A_2A_3}M^{B_1B_2}\,\!_E &=&
-{10 \over 9}\bigl({\cal{M}}_{12}\,\!^{S;B_1B_2A_1A_2A_3}
  +{\cal{M}}_{12}^{B_1;B_2SA_1A_2A_3}\bigr)  \\
  &+&{2 \over 27}\bigl(8\eta^{A_1B_1}{\cal{M}}_{12}\,\!^{A_2A_3;B_2S}
  +\eta^{SA_1}{\cal{M}}_{12}\,\!^{A_2A_3;B_1B_2}\bigr) \\
\end{eqnarray*}
\begin{eqnarray*}
{\cal{M}}_{10}\,\!^{S_1S_2;A_1A_2E}M^{B_1B_2}\,\!_E &=&
  -{10 \over 9}{\cal{M}}_{12}\,\!^{S_1;S_2A_1A_2B_1B_2} \\
  &+&{4 \over 7\times 81}\bigl\{58\eta^{A_1B_1}{\cal{M}}_{12}^{S_1B_2;S_2A_2}
  +31\eta^{S_1B_1}{\cal{M}}_{12}\,\!^{A_1A_2;S_2B_2} \\
  &+&11\eta^{S_1A_1}{\cal{M}}_{12}^{A_2S_2;B_1B_2}
  -2\eta^{S_1S_2}{\cal{M}}_{12}^{A_1A_2;B_1B_2}\bigr\} \\
\end{eqnarray*}

\begin{eqnarray}
\lefteqn{{\cal{M}}_{10}\,\!^{S_1S_2;A_1A_2A_3} M^{B_1B_2B_3}
  =\hat{\cal{M}}_{10}^{S_1S_2[A_1;A_2A_3]}M^{B_1B_2B_3} }\nonumber \\
  & &=-{15\over 11}\biggl\{{6\over 7}\bigl(\eta^{S_1B_1}
  {\cal{M}}_{12}^{S_2;A_1A_2A_3B_2B_3}+\eta^{S_1B_1}
  {\cal{M}}_{12}^{A_1;S_2A_2A_3B_2B_3}+\eta^{A_1B_1}
  {\cal{M}}_{12}^{S_1;S_2A_2A_3B_2B_3}\bigr) \nonumber \\
  & &-{26\over 63}\eta^{S_1A_1}{\cal{M}}_{12}^{S_2;A_2A_3B_1B_2B_3}
  -{8\over 63}\eta^{S_1A_1}{\cal{M}}_{12}^{A_2;S_2A_3B_1B_2B_3}
  -{1\over 7}\eta^{S_1S_2}{\cal{M}}_{12}^{A_1;A_2A_3B_1B_2B_3}\biggr\}
\nonumber \\
  & &+{2 \over 45}\eta^{S_1A_1}\eta^{S_2B_1}{\cal{M}}_{12}^{A_2A_3;B_2B_3}
  -{32\over 21\times 15}\eta^{S_1A_1}\eta^{A_2B_1}
  {\cal{M}}_{12}^{B_2B_3;A_3S_2}\nonumber \\
  & &+{12\over 35}\eta^{S_1B_1}\eta^{A_1B_2}{\cal{M}}_{12}^{A_2A_3;B_3S_2}
+{8\over 35}\eta^{A_1B_1}\eta^{A_2B_2}{\cal{M}}_{12}^{S_1A_3;S_2B_3}
\nonumber \\
  & &-{1\over 35}\eta^{S_1S_2}\eta^{A_1B_1}
  {\cal{M}}_{12}^{A_2A_3;B_2B_3}
\end{eqnarray}

\no The second irreducible piece has 7 indices; we can extract a seven-index
object by removing one of the factors from $\hat{\cal{M}}_{12}^{XABY;C_1C_2}$
to obtain the structure

\be
\hat{\cal{M}}_{10}^{XYZ;A_1A_2B_1B_2}=M^{XED_1}M^{Y}\,\!_{D_1D_2}
  M^{ZD_2F}M^{A_1A_2}\,\!_EM_F\,\!^{B_1B_2}.
\ee

\no It is clear that

\[\hat{\cal{M}}_{10}^{ ^{\scriptstyle [X} YZ; ^{\scriptstyle A_1A_2]} B_1B_2}=
0
\qquad \hat{\cal{M}}_{10}^{XY ^{\scriptstyle [Z}
;A_1A_2 \,\!^{\scriptstyle B_1B_2]} }= 0 \]
\be
\hat{\cal{M}}_{10}^{XYZ;A_1A_2B_1B_2}=
- \hat{\cal{M}}_{10}^{ZYX;B_1B_2A_1A_2}
\ee

\no aside from the obvious antisymmetry in $A_{1}, A_{2}$ and $B_{1}, B_{2}$.

        The object in (4.50) is not irreducible because it is not traceless.
Its only non-vanishing traces are

\[\hat{\cal M}_{10} \,_{E} \,\!^{YZ;A_1A_2 \, EB} =
- \hat{\cal M}_{10}^{YZB;A_1A_2}\]
\[\hat{\cal M}_{10}^{XY} \,\!_{E} \,\!^{;EA \, B_1B_2}  =
\hat{\cal M}_{10}^{YXA;B_1B_2}\]
\[\hat{\cal M}_{10}^{XYZ;} \,\!_{E} \,\!^{A \, EB}  =
- {\cal N}^{YXZBA}\]
\be
{\cal N}^{YXZBA} =:
M^{XD_1} \,\!_{D_2} M^{YD_2} \,\!_{D_3} M^{ZD_3} \,\!_{D_4} M^{BD_4} \,\!_{D_5}
 M^{AD_5} \,\!_{D_1}
\ee

For ${\cal N}^{XYZBA}$ we have

\[ {\cal N}^{XYZBA} =
{\cal N}^{AXYZB} \]
\be
{\cal N}^{XYZBA} =
- {\cal N}^{ABZYX}
\ee

\no as well as, by using (\ref{friend}) in the last two factors,

\be
{\cal N}^{XYZAB} =
{\cal N}^{XYZBA} + \hat{\cal M}_{10}^{XYZ;BA}
\ee

Iterating (4.54) and using (4.44) one can derive the decomposition

\be
{\cal N}^{XYZAB} =
- \frac{1}{2} \hat{\cal M}_{10}^{ZAB;XY}
- \hat{\cal M}_{10}^{BA(X;Y)Z}
+ \frac{1}{2} \hat{\cal M}_{10}^{XYZ;BA}
\ee

\no and therefore

\begin{eqnarray}
{\cal N}^{[XY]ZAB} &=&
- \frac{1}{2} \hat{\cal M}_{10}^{ZAB;XY}
\nonumber \\
{\cal N}^{[X^{\scriptstyle Y} Z]AB} &=&
- \hat{\cal M}_{10}^{AB[X;Z]Y}
- \frac{1}{2} \hat{\cal M}_{10}^{ABY;XZ}
\end{eqnarray}

\no expressions that will be needed later.

In order to obtain the second irreducible piece of this $\theta^{10}$ sector we
can just project (4.50) according to the pattern
\setlength{\unitlength}{.5pt}
$\begin{picture}(20, 30)
\multiput(0, 20)(0, -10){5}{\framebox(10,10){ }}
\multiput(10, 20)(0, -10){2}{\framebox(10,10){ }}
\end{picture}$. One obtains the structure

\begin{eqnarray}
{\cal M}_{10}^{CD;A_1 \ldots A_5} &=& \hat{\cal M}_{10}^{CA_1D;A_2 \ldots A_5}
\nonumber \\
  &=&M^{CEG_1} M^{A_1}\,\!_{G_1G_2}M^{DG_2F}M^{A_2A_3}\,\!_{E}
M_F \,\!^{A_4A_5},
\end{eqnarray}

\no which is completely antisymmetric in $A_{1},...,A_{5}$
(we remind the reader of
our letter convention) and by virtue of (4.51) it is also antisymmetric in
$C, D$:

\be
{\cal M}_{10}^{CD;A_1 \ldots A_5} = - {\cal M}_{10}^{DC;A_1 \ldots A_5}
\ee

\no Its tracelessness is immediate from (\ref{nueva}), (\ref{friend}),
(2.3) and (4.3), and it also satisfies

\be
{\cal M}_{10}^{C[D;A_1 \ldots A_5]} = 0
\qquad {\cal M}_{10}^{[C_1C_2;B_1 \ldots B_4]A} = 0
\ee

\no and it is self-dual:

\be
{\cal{M}}_{10}^{C_1C_2;B_1\ldots B_5}= {1 \over 5!}\epsilon^{B_1\ldots B_5
  D_1\ldots D_5} {\cal{M}}_{10}\,\!^{C_1C_2;}\,\!_{D_1\ldots D_5}
\ee

The list of decompositions is:

\[{\cal{M}}_{10}^{A_1A_2;B_1B_2E_1E_2E_3}M_{E_1E_2E_3} =
  \frac{2}{5}{\cal{M}}_{12}\,\!^{A_1A_2;B_1B_2} \]

\begin{eqnarray*}
{\cal M}_{10}^{E_1A;B_1B_2B_3E_2E_3}M_{E_1E_2E_3} = 0&
&{\cal{M}}_{10}^{E_1E_2;B_1 \ldots B_4E_3}M_{E_1E_2E_3} = 0 \\
{\cal M}_{10}^{E_1A_;B_1 \ldots B_4E_2}M^C \,\!_{E_1E_2} =
{\cal M}_{12}^{[C;A]B_1 \ldots B_4}&
&{\cal{M}}_{10}^{E_1E_2;B_1 \ldots B_5}M^C \,\!_{E_1E_2} =
2 {\cal M}_{12}^{C;B_1 \ldots B_5}
\end{eqnarray*}

\begin{eqnarray*}
\lefteqn{{\cal{M}}_{10}^{A_1A_2;B_1B_2B_3E_1E_2}M^C \,\!_{E_1E_2} = } \\
& &{\cal M}_{12}^{[C;A_1A_2]B_1B_2B_3}
+ \frac{2}{45} \eta^{A_1B_1} {\cal M}_{12}^{A_2C;B_2B_3}
+ \frac{7}{45} \eta^{CB_1} {\cal M}_{12}^{A_1A_2;B_2B_3}
\end{eqnarray*}

\begin{eqnarray*}
\lefteqn{{\cal{M}}_{10}^{A_1A_2;B_1 \ldots B_4E}M^{C_1C_2} \,\!_E =
\frac{1}{11} \biggl[\frac{5}{2} \eta^{A_1C_1} \left(
{\cal M}_{12}^{A_2;C_2B_1 \ldots B_4}
- {\cal M}_{12}^{C_2;A_2B_1 \ldots B_4} \right)} \\
& &+\frac{2}{3}\eta^{A_1B_1}\left(7{\cal M}_{12}^{C_1;C_2A_2B_2B_3B_4}
+ 5{\cal M}_{12}^{A_2;C_1C_2B_2B_3B_4}\right) \\
& &+2\eta^{B_1C_1}\left(5{\cal M}_{12}^{A_1;A_2C_2B_2B_3B_4}
+ 2{\cal M}_{12}^{C_2;A_1A_2B_2B_3B_4}\right) \biggr] \\
& &+\frac{1}{9 \times 25}\left(\eta^{A_1B_1}\eta^{A_2B_2}
{\cal M}_{12}^{B_3B_4;C_1C_2}
+21 \eta^{B_1C_1}\eta^{B_2C_2} {\cal M}_{12}^{A_1A_2;B_3B_4}
-14 \eta^{A_1B_1}\eta^{C_1B_2} {\cal M}_{12}^{A_2C_2;B_3B_4} \right)
\end{eqnarray*}

\begin{eqnarray*}
{\cal{M}}_{10}^{EA;B_1 \ldots B_5}M^{C_1C_2} \,\!_E &=&
\frac{5}{11} \biggl[ \eta^{AC_1} {\cal M}_{12}^{C_2;B_1 \ldots B_5}
+ \frac{3}{2} \eta^{AB_1}
{\cal M}_{12}^{C_1;C_2B_2 \ldots B_5} \\
& &+ \frac{3}{2} \eta^{B_1C_1}
{\cal M}_{12}^{C_2;AB_2 \ldots B_5}
- \frac{5}{2} \eta^{B_1C_1}
{\cal M}_{12}^{A;C_2B_2 \ldots B_5} \biggr]
\end{eqnarray*}

\begin{eqnarray}
\lefteqn{{\cal{M}}_{10}^{A_1A_2;B_1 \ldots B_5}M^{C_1C_2C_3} =} \nonumber \\
& &-\frac{1}{6!} \left( \frac{1}{10} \epsilon^{B_1 \ldots B_5A_1A_2C_3E_1E_2}
{\cal M}_{12}^{C_1C_2;} \,\!_{E_1E_2} +
\frac{1}{2} \epsilon^{B_1 \ldots B_5C_1C_2C_3E_1E_2}
{\cal M}_{12}^{A_1A_2;} \,\!_{E_1E_2} \right. \nonumber \\
& &\left. - \frac{3}{5} \epsilon^{B_1 \ldots B_5A_2C_2C_3E_1E_2}
{\cal M}_{12}^{A_1C_1;} \,\!_{E_1E_2} \right) \nonumber \\
& &+\frac{1}{11} \bigl[ 2 \eta^{A_1C_1} \eta^{A_2C_2}
{\cal M}_{12}^{C_3;B_1 \ldots B_5}
+5 \eta^{A_1B_1} \eta^{A_2B_2}
{\cal M}_{12}^{C_1;C_2C_3B_3B_4B_5}  \nonumber \\
& &-\frac{15}{2} \eta^{A_1B_1} \eta^{A_2C_1}
{\cal M}_{12}^{C_2;C_3B_2 \ldots B_5}
- 15 \eta^{B_1A_1} \eta^{B_2C_1} \left(
{\cal M}_{12}^{A_2;C_2C_3B_3B_4B_5}
+ {\cal M}_{12}^{C_2;C_3A_2B_3B_4B_5} \right) \nonumber \\
& &+ 5 \eta^{B_1C_1} \eta^{B_2C_1} \left(
3 {\cal M}_{12}^{A_1;A_2C_3B_3B_4B_5}
+ {\cal M}_{12}^{C_3;A_1A_2B_3B_4B_5} \right) \nonumber \\
& & + \frac{5}{4} \eta^{C_1A_1} \eta^{C_2B_1} \left(
5 {\cal M}_{12}^{C_3;A_2B_2 \ldots B_5}
- 9 {\cal M}_{12}^{A_2;C_3B_2 \ldots B_5} \right) \bigr] \nonumber \\
& &+ \frac{1}{12} \eta^{B_1C_1} \eta^{B_2C_2}\eta^{B_3C_3}
{\cal M}_{12}^{A_1A_2;B_4B_5}
+ \frac{1}{10} \eta^{B_1A_1} \eta^{B_2C_1}\eta^{B_3C_2}
{\cal M}_{12}^{A_2C_3;B_4B_5} \nonumber \\
& &+ \frac{1}{60} \eta^{B_1A_1} \eta^{B_2A_2}\eta^{B_3C_1}
{\cal M}_{12}^{C_2C_3;B_4B_5}
\end{eqnarray}

\ve
\no$\underline{\theta^{8}}$.

At the beginning of this section we introduced one of the irreducible parts of
the $\theta^{8}$ sector, namely the totally symmetric tensor in (4.1):

\be
{\cal{M}}_8^{S_1S_2S_3S_4}=M^{S_1}\,\!_E\,\!^FM^{S_2}\,\!_F\,\!^G
M^{S_3}\,\!_G\,\!^H M^{S_4}\,\!_H\,\!^E.
\ee

\no Its products with $M^{A_{1} A_{2} A_{3}}$ are particularly easy to
decompose using (4.43):

\[ {\cal{M}}_8^{S_1S_2S_3E} M^{A_1A_2} \,\!_E =
\hat{\cal M}_{10} ^{S_1S_2S_3;A_1A_2} \]
\begin{eqnarray}
{\cal{M}}_8^{S_1S_2S_3S_4} M^{A_1A_2A_3} &=&
\frac{7}{6} \eta^{S_1A_1} \hat{\cal M}_{10} ^{S_2S_3S_4;A_2A_3}
-\frac{1}{4} \eta^{S_1S_2} \hat{\cal M}_{10} ^{S_3S_4A_1;A_2A_3}
\nonumber \\
&=&\frac{21}{10} \eta^{S_1A_1} {\cal M}_{10} ^{S_2S_3;S_4A_2A_3}
-\frac{1}{4} \eta^{S_1S_2} {\cal M}_{10} ^{S_3S_4;A_1A_2A_3}
\end{eqnarray}

This sector contains two additional irreducible pieces (see Table 1). In order
to isolate them, first we remove one factor from
$\hat{\cal M}_{10}^{S_1S_2S_3;A_1A_2}$ to get the structure

\be
\hat{\cal{M}}_8^{XY\, A_1A_2\, B_1B_2}=
M^{XED}M^Y\,\!_D\,\!^FM^{A_1A_2}\,\!_E M^{B_1B_2}\,\!_F
\ee

\no with the following properties

\[ \hat{\cal{M}}_8^{XY\, A_1A_2\, B_1B_2}=
\hat{\cal{M}}_8^{YX\, B_1B_2\, A_1A_2} \]
\be
\hat{\cal{M}}_{8}^{[X ^{\scriptstyle Y} \, A_1A_2] \, B_1B_2 }= 0
\qquad \hat{\cal{M}}_{8}^{X[Y \, ^{\scriptstyle A_1A_2} \, B_1B_2] }= 0
\ee

\no It is reducible,

\be
\hat{\cal M}_8^{XY \, EA}\,\!_{E} \,\!^B =
-{\cal M}_8^{XYAB}
\ee

\no but easy to detrace:

\be
\hat{\cal M}_8^{XY \, A_1A_2\, B_1B_2} =
Traceless(\hat{\cal M}_8^{XY \, A_1A_2\, B_1B_2})
- \frac{1}{2} \eta^{A_1B_1} {\cal M}_8^{XYA_2B_2}
\ee

The traceless part is going to contain 2 irreducible pieces corresponding to
the patterns
\setlength{\unitlength}{.5pt}
$\begin{picture}(30, 40)
\multiput(0, 10)(0, -10){4}{\framebox(10,10){ }}
\multiput(10, 10)(10, 0){2}{\framebox(10,10){ }}
\end{picture}$  and
\setlength{\unitlength}{.5pt}
$\begin{picture}(20, 30)
\multiput(0, 0)(10, 0){2}{\framebox(10,10){ }}
\multiput(0, 10)(10, 0){2}{\framebox(10,10){ }}
\multiput(0, -10)(10, 0){2}{\framebox(10,10){ }}
\end{picture}$ . First,

\be
Y\left(\setlength{\unitlength}{.5pt}
\begin{picture}(30, 30)
\multiput(0, 15)(0, -10){4}{\framebox(10,10){ }}
\multiput(10, 15)(10, 0){2}{\framebox(10,10){ }}
\end{picture}
\right) \hat{\cal M}_8^{XY \, A_1A_2\, B_1B_2} =
\hat{\cal M}_8^{XY \, [A_1A_2\, B_1B_2]} +
2 \hat{\cal M}_8^{A_1B_1 \, [XA_2\, YB_2]}
\ee

\no and thus the
\setlength{\unitlength}{.5pt}
$\begin{picture}(30, 30)
\multiput(0, 10)(0, -10){4}{\framebox(10,10){ }}
\multiput(10, 10)(10, 0){2}{\framebox(10,10){ }}
\end{picture}$
irreducible structure is

\be
{\cal M}_8^{XY;B_1B_2B_3B_4} =
\hat{\cal M}_8^{XY \, B_1B_2\, B_3B_4}
\ee

\no which is completely antisymmetric in $B_{1},...,B_{4}$ and, by (4.65),
symmetric in $X, Y$:

\be
{\cal M}_8^{XY;B_1B_2B_3B_4} =
{\cal M}_8^{YX;B_1B_2B_3B_4}
\ee

\no and satisfying

\be
{\cal M}_8^{X[Y;B_1B_2B_3B_4]} = 0.
\ee

Second,

\begin{eqnarray}
Y\left(
\setlength{\unitlength}{.5pt}
\begin{picture}(20, 30)
\multiput(0, 0)(10, 0){2}{\framebox(10,10){ }}
\multiput(0, 10)(10, 0){2}{\framebox(10,10){ }}
\multiput(0, -10)(10, 0){2}{\framebox(10,10){ }}
\end{picture}
\right)\hat{\cal M}_8^{XY \, A_1A_2\, B_1B_2} &=&
- \frac{1}{2} \left(
\hat{\cal{M}}_{8}^{[X ^{\scriptstyle [A_1\,B_1B_2]} YA_2] }
+ \hat{\cal{M}}_{8}^{[X ^{\scriptstyle [A_1\,YA_2]} B_1B_2] } \right.
\nonumber \\
&+& \left. \hat{\cal{M}}_{8}^{[A_1 ^{\scriptstyle [X\,B_1A_2]} YB_2] }
+ \hat{\cal{M}}_{8}^{[X ^{\scriptstyle [A_1\,B_1A_2]} YB_2] } \right)
\end{eqnarray}

\no giving as the
\setlength{\unitlength}{.5pt}
$\begin{picture}(20, 30)
\multiput(0, 0)(10, 0){2}{\framebox(10,10){ }}
\multiput(0, 10)(10, 0){2}{\framebox(10,10){ }}
\multiput(0, -10)(10, 0){2}{\framebox(10,10){ }}
\end{picture}$ irreducible structure the object

\be
{\cal M}_8^{A_1A_2A_3; B_1B_2B_3} =
\hat{\cal M}_8^{A_1B_1 \, B_2B_3\, A_2A_3}
\ee

Of course it is completely antisymmetric in the $A$ and $B$ indices separately
and, from (4.65), we see that it is symmetric upon interchange of both groups
of indices

\be
{\cal M}_8^{A_1A_2A_3; B_1B_2B_3} =
{\cal M}_8^{B_1B_2B_3; A_1A_2A_3}
\ee

The remaining important property of this tensor can be derived from the
definitions (4.73), (4.64) by using once more the properties of the
$\theta^{4}$ sector,

\be
{\cal M}_8^{A_1A_2[C; B_1B_2B_3]} = 0
\ee

\no that implies also

\be
{\cal M}_8^{A[B_1B_2;B_3B_4]C} = 0
\ee

By using the properties in (4.72) we can write finally for the decomposition
in (4.67):

\begin{eqnarray}
\hat{\cal M}_8^{XY \, A_1A_2\, B_1B_2} &=&
{\cal M}_8^{XY;A_1A_2B_1B_2}
+ 2{\cal M}_8^{A_1B_1;XA_2YB_2} \nonumber \\
&+& \frac{3}{8} \left(
3 {\cal M}_8^{YA_1A_2; XB_1B_2} - {\cal M}_8^{XA_1A_2; YB_1B_2} \right)
\nonumber \\
&-&\frac{1}{2} \eta^{A_1B_1} {\cal M}_8^{XYA_2B_2}
\end{eqnarray}

The lists of products decompositions for these irreducible pieces are

\[{\cal{M}}_{10}^{A_1A_2;B_1B_2E_1E_2E_3}M_{E_1E_2E_3} =
  \frac{2}{5}{\cal{M}}_{12}\,\!^{A_1A_2;B_1B_2} \]

\begin{eqnarray*}
{\cal M}_{8}^{S_1S_2;BE_1E_2E_3}M_{E_1E_2E_3} = 0&
&{\cal M}_{8}^{SE_1;B_1B_2E_2E_3}M_{E_1E_2E_3} = 0 \\
{\cal M}_{8}^{SE_1;B_1B_2B_3E_2}M^C \,\!_{E_1E_2} =
-\frac{1}{2} {\cal M}_{10}^{SC;B_1B_2B_3} &
&{\cal M}_{8}^{E_1E_2;B_1 \ldots B_4}M^C \,\!_{E_1E_2} = 0
\end{eqnarray*}

\begin{eqnarray*}
{\cal{M}}_{8}^{S_1S_2;B_1B_2E_1E_2}M^C \,\!_{E_1E_2} &=&
\frac{1}{3} \left(2 \hat{\cal M}_{10}^{S_1S_2B_1;B_2C}
- \hat{\cal M}_{10}^{S_1S_2C;B_1B_2} \right) \\
&=&\frac{1}{5} \left(3 {\cal M}_{10}^{S_1S_2;B_1B_2C}
- \hat{\cal M}_{10}^{CS_1;S_2B_1B_2} \right)
\end{eqnarray*}

\begin{eqnarray*}
{\cal{M}}_{8}^{S_1S_2;B_1B_2B_3E}M^{C_1C_2} \,\!_E &=&
\frac{10}{9} {\cal M}_{10}^{S_1C_1;S_2C_2B_1B_2B_3} \\
&+& \frac{1}{6} \left( \eta^{S_1C_1} \hat{\cal M}_{10}^{S_2C_2B_1;B_2B_3}
+ \eta^{S_1B_1} \hat{\cal M}_{10}^{S_2C_1B_2;B_3C_2} \right) \\
&+& \frac{1}{4} \eta^{B_1C_1}\left(-\frac{7}{3}
\hat{\cal M}_{10}^{S_1S_2B_2;B_3C_2}
+ \hat{\cal M}_{10}^{S_1S_2C_2;B_2B_3} \right) \\
&=&\frac{10}{9} {\cal M}_{10}^{S_1C_1;S_2C_2B_1B_2B_3} \\
&+& \frac{1}{6} \eta^{S_1C_1} {\cal M}_{10}^{S_2C_2;B_1B_2B_3}
+ \frac{1}{4} \eta^{S_1B_1} {\cal M}_{10}^{S_2C_1;B_2B_3C_2} \\
&+& \frac{1}{20} \eta^{B_1C_1}\left(-11 {\cal M}_{10}^{S_1S_2;B_2B_3C_2}
+ 13 {\cal M}_{10}^{S_1C_2;S_2B_2B_3} \right)
\end{eqnarray*}

\begin{eqnarray*}
{\cal{M}}_{8}^{SE;B_1 \ldots B_4}M^{C_1C_2} \,\!_E &=&
\frac{5}{9} \left( {\cal M}_{10}^{C_1C_2;SB_1 \ldots B_4}
+ 5 {\cal M}_{10}^{SC_1;C_2B_1 \ldots B_4} \right) \\
&+& \frac{2}{3} \eta^{B_1C_1} {\cal M}_{10}^{SC_2;B_2B_3B_4}
\end{eqnarray*}

\begin{eqnarray}
\lefteqn{{\cal{M}}_{8}^{S_1S_2;B_1 \ldots B_4}M^{C_1C_2C_3} =
-\frac{2}{3 \times 5!} \epsilon^{B_1 \ldots B_4^{\scriptstyle [C_1}
C_2C_3E_1E_2E_3} {\cal M}_{10}^{ ^{\scriptstyle S_1]} S_2;} \,\!_{E_1E_2E_3} }
\nonumber \\
& &+ \frac{4}{21} \left[
\eta^{S_1C_1} \left( 10 {\cal M}_{10}^{S_2C_2;C_3B_1 \ldots B_4}
+ 2 {\cal M}_{10}^{C_2C_3;S_2B_1 \ldots B_4} \right)  \right.
\nonumber \\
& & - \eta^{S_1B_1} \left( 6 {\cal M}_{10}^{S_2C_1;C_2C_3B_2B_3B_4}
+ {\cal M}_{10}^{C_1C_2;S_2C_3B_2B_3B_4} \right)
\nonumber \\
& & \left.- 8 \eta^{B_1C_1} {\cal M}_{10}^{S_1C_2;S_2C_3B_2B_3B_4}
- \eta^{S_1S_2} {\cal M}_{10}^{C_1C_2;C_3B_1 \ldots B_4} \right]
\nonumber \\
& &+ \frac{1}{5} \left[
2 \eta^{S_1C_1} \eta^{B_1C_2} {\cal M}_{10}^{S_2C_3;B_2B_3B_4}
+ 3 \eta^{S_1B_1} \eta^{C_1B_2} {\cal M}_{10}^{S_2C_2;C_3B_3B_4} \right.
\nonumber \\
& &+3 \eta^{B_1C_1}\eta^{B_2C_2} \left. \left( {\cal M}_{10}^{S_1S_2;C_3B_3B_4}
- {\cal M}_{10}^{S_1C_3;S_2B_3B_4} \right) \right]
\end{eqnarray}

\no and

\begin{eqnarray*}
{\cal M}_{8}^{A_1A_2A_3;E_1E_2E_3}M_{E_1E_2E_3} = 0&
&{\cal M}_{8}^{A_1A_2E_1;BE_2E_3}M_{E_1E_2E_3} = 0 \\
{\cal M}_{8}^{A_1A_2A_3;BE_1E_2}M^C \,\!_{E_1E_2} =
\frac{2}{3} {\cal M}_{10}^{BC;A_1A_2A_3} &
&{\cal M}_{8}^{A_1A_2E_1;B_1B_2E_2}M^C \,\!_{E_1E_2} =
-\frac{2}{3} {\cal M}_{10}^{CA_1;A_2B_1B_2}
\end{eqnarray*}

\[{\cal{M}}_{8}^{A_1A_2E;B_1B_2B_3}M^{B_4B_5} \,\!_E =
- \frac{2}{3} {\cal M}_{10}^{A_1A_2;B_1 \ldots B_5} \]

\begin{eqnarray*}
\lefteqn{
{\cal{M}}_{8}^{A_1A_2A_3;B_1B_2E}M^{C_1C_2} \,\!_E =} \\
& &-\frac{8}{9} \left( \frac{2}{3} {\cal M}_{10}^{B_1B_2;A_1A_2A_3C_1C_2}
+ {\cal M}_{10}^{A_1A_2;A_3B_1B_2C_1C_2}
-\frac{1}{6} {\cal M}_{10}^{C_1C_2;A_1A_2A_3B_1B_2} \right) \\
& & \frac{1}{15} \left[\frac{7}{3} \eta^{B_1C_1}
{\cal M}_{10}^{B_2C_2;A_1A_2A_3}
+ \frac{7}{2} \eta^{A_1C_1} {\cal M}_{10}^{C_2A_2;A_3B_1B_2}
- \frac{3}{2} \eta^{A_1B_1} {\cal M}_{10}^{A_2B_2;A_3C_1C_2} \right]
\end{eqnarray*}

\begin{eqnarray}
\lefteqn{{\cal{M}}_{8}^{A_1A_2A_3;B_1B_2B_3}M^{C_1C_2C_3} = }
\nonumber \\
&+& \frac{4}{15} \left\{ \eta^{B_3C_3} \left(
-4 {\cal M}_{10}^{B_1B_2;A_1A_2A_3C_1C_2}
+ {\cal M}_{10}^{C_1C_2;A_1A_2A_3B_1B_2}
-6 {\cal M}_{10}^{A_1A_2;A_3B_1B_2C_1C_2}
\right) \right. \nonumber \\
& &+\eta^{A_3C_3} \left(
-4 {\cal M}_{10}^{A_1A_2;B_1B_2B_3C_1C_2}
+ {\cal M}_{10}^{C_1C_2;B_1B_2B_3A_1A_2}
-6 {\cal M}_{10}^{B_1B_2;B_3A_1A_2C_1C_2}
\right)
\nonumber \\
& &\left. +3\eta^{A_3B_3} \left(
{\cal M}_{10}^{A_1A_2;B_1B_2C_1C_2C_3}
+{\cal M}_{10}^{B_1B_2;A_1A_2C_1C_2C_3}
- {\cal M}_{10}^{C_1C_2;C_3A_1A_2B_1B_2}
\right) \right\}
\nonumber \\
&+& \frac{14}{75} \bigl[
\eta^{B_1C_1} \eta^{B_2C_2} {\cal M}_{10}^{B_3C_3;A_1A_2A_3}
+ \eta^{A_1C_1} \eta^{A_2C_2} {\cal M}_{10}^{A_3C_3;B_1B_2B_3}
\nonumber \\
& &-\frac{3}{2} \eta^{B_1C_1} \eta^{B_2A_1} {\cal M}_{10}^{B_3A_2;A_3C_2C_3}
-\frac{3}{2} \eta^{A_1C_1} \eta^{A_2B_1} {\cal M}_{10}^{A_3B_2;B_3C_2C_3}
\nonumber \\
& & - 3 \eta^{A_1C_1} \eta^{B_1C_2} {\cal M}_{10}^{C_3A_2;A_3B_2B_3}
+ \frac{1}{7} \eta^{A_1B_1} \eta^{A_2B_2} {\cal M}_{10}^{A_3B_3;C_1C_2C_3}
\bigr]
\end{eqnarray}

\no$\underline{\theta^{6}}$.

In the decomposition (3.10) of the product of three $M^{A_{1} A_{2} A_{3}}$
we have two types of irreducible structures:

\be
\hat{\cal{M}}_6^{A\, B_1B_2\, C_1C_2}=
M^{ADE}M^{B_1B_2}\,\!_DM^{C_1C_2}\,\!_E
\ee

\no and

\be
{\cal{M}}_6^{A_1A_2; B_1 \ldots B_5}=
M^{A_1A_2E}M_E \,\!^{B_1B_2}M^{B_3B_4B_5}
\ee

The expression (4.80) trivially satisfies

\be
\hat{\cal{M}}_6^{A\, B_1B_2\, C_1C_2}=
- \hat{\cal{M}}_6^{A\, C_1C_2\,  B_1B_2}
\ee

\no and (\ref{friend}) implies

\be
\hat{\cal{M}}_6^{[A\, B_1B_2]\, C_1C_2}=0
\ee

The tensor ${\cal M}_{6}^{[A B_{1} B_{2}] C_{1} C_{2}}$ must belong to the
representation
\begin{picture}(30, 30)
\multiput(0, 10)(0, -10){3}{\framebox(10,10){ }}
\multiput(10, 10)(10, 0){2}{\framebox(10,10){ }}
\end{picture}
and in order to make the corresponding Young symmetry
obvious, we define the new tensor

\be
{\cal{M}}_6^{XY;B_1B_2B_3}=
\hat{\cal{M}}_6^{X\, YB_1 \, B_2B_3}
\ee

Both tensors are completely equivalent though, the inverse of (4.84) being

\be
\hat{\cal{M}}_6^{A\, B_1B_2\, C_1C_2}=
3 {\cal{M}}_6^{A B_1;B_2C_1C_2}
\ee

It is easy to see that ${\cal M}_{6}^{X Y; B_{1} B_{2} B_{3}}$ must be
symmetric
in $X, Y$:

\begin{eqnarray}
\lefteqn{{\cal M}_6^{[XY];B_1B_2B_3} =M^{DE[X}M^{Y]B_1}\,\!_D M^{B_2B_3}\,\!_E
=}
\nonumber \\
& &-\frac{1}{2}M^{XY} \,\!_DM^{DE[B_1} M^{B_2B_3]} \,\!_E =0
\end{eqnarray}

\no where we have used (\ref{friend}) twice. Thus

\be
{\cal{M}}_6^{XY;B_1B_2B_3}=
{\cal{M}}_6^{YX;B_1B_2B_3}
\ee

\no The remaining important property of this tensor is

\be
{\cal{M}}_6^{X[Y;B_1B_2B_3]}=0
\ee

\no as we have come to expect and can be immediately seen from (4.84) and
(4.80). This time we have the following product decompositions:

\begin{eqnarray*}
{\cal M}_{6}^{S_1S_2;E_1E_2E_3}M_{E_1E_2E_3} = 0&
&{\cal M}_{6}^{SE_1;BE_2E_3}M_{E_1E_2E_3} = 0 \\
{\cal M}_{6}^{SE_1;B_1B_2E_2}M^C \,\!_{E_1E_2} = 0 &
&{\cal{M}}_{6}^{S_1S_2;BE_1E_2}M^C \,\!_{E_1E_2} =
- \frac{2}{3} {\cal M}_{8}^{S_1S_2BC}
\end{eqnarray*}

\begin{eqnarray*}
{\cal{M}}_{6}^{S_1S_2;B_1B_2E}M^{C_1C_2} \,\!_E &=&
- \frac{4}{3} {\cal M}_{8}^{S_1C_1;S_2C_2B_1B_2}
+ \frac{2}{3} {\cal M}_{8}^{S_1S_2;B_1B_2C_1C_2} \\
&-& \frac{1}{2} {\cal M}_{8}^{S_1B_1B_2;S_2C_1C_2}
- \frac{1}{3} \eta^{B_1C_1} {\cal M}_{8}^{S_1S_2B_2C_2} \\
{\cal{M}}_{6}^{SE;B_1B_2B_3}M^{C_1C_2} \,\!_E &=&
\frac{4}{3} {\cal M}_{8}^{SC_1;C_2B_1B_2B_3}
-{\cal M}_{8}^{SC_1C_2;B_1B_2B_3}
\end{eqnarray*}

\[{\cal{M}}_{6}^{S_1S_2;B_1B_2B_3}M^{B_4B_5B_6} =
 \frac{1}{2 \times 5!} \epsilon^{B_1 \ldots B_6 E_1 \ldots E_4}
{\cal M}_{8}^{S_1S_2;} \,\!_{E_1 \ldots E_4} \]

\begin{eqnarray}
\lefteqn{{\cal{M}}_{6}^{S_1S_2;B_1B_2B_3}M^{C_1C_2C_3} = } \nonumber \\
& &\frac{3}{8 \times 5!} \left(
\epsilon^{B_1B_2B_3C_1C_2C_3E_1 \ldots E_4}
{\cal M}_{8}^{S_1S_2;} \,\!_{E_1 \ldots E_4}
+ 2\epsilon^{S_1B_1B_2C_1C_2C_3E_1 \ldots E_4}
{\cal M}_{8}^{S_2B_3;} \,\!_{E_1 \ldots E_4} \right) \nonumber \\
& & +9 \eta^{B_1C_1} \left( -\frac{1}{5} {\cal M}_{8}^{S_1C_2;S_2C_3B_2B_3}
+ {\cal M}_{8}^{S_1S_2;B_2B_3C_2C_3} \right)
\nonumber \\
& & +\frac{9}{10} \eta^{S_1C_1} {\cal M}_{8}^{S_2C_2;C_3B_1B_2B_3}
+\frac{1}{2} \eta^{S_1B_1} {\cal M}_{8}^{S_2B_2;B_3C_1C_2C_3}
\nonumber \\
& & + \frac{3}{56} \left[
-9 \eta^{B_1C_1} {\cal M}_{8}^{S_1B_2B_3;S_2C_2C_3}
- 12 \eta^{S_1C_1} {\cal M}_{8}^{S_2C_2C_3;B_1B_2B_3} \right.
\nonumber \\
& &\left. +2 \eta^{S_1B_1} {\cal M}_{8}^{S_2B_2B_3;C_1C_2C_3}
+ \eta^{S_1S_2} {\cal M}_{8}^{B_1B_2B_3;C_1C_2C_3} \right]
\nonumber \\
& &-\frac{3}{14} \eta^{B_1C_1} \eta^{B_2C_2} {\cal M}_{8}^{S_1S_2B_3C_3}
\end{eqnarray}

Turning our attention to (4.81), we get the duality property

\be
{\cal{M}}_6^{A_1A_2;B_1\ldots B_5}=-{1 \over 5!}\epsilon^{B_1\ldots B_5
  D_1\ldots D_5} {\cal{M}}_6\,\!^{A_1A_2;}\,\!_{D_1\ldots D_5}
\ee

\no as a direct consequence of the one for ${\cal M}_{4}^{C; D_{1}...D_{5}}$
(eq. (3.8)). The following bracket property is also immediate

\be
{\cal M}_{6}^{A[C;B_1 \ldots B_5]} = 0
\ee

Finally, to complete this section we have the following list of decompositions:

\[{\cal M}_{6}^{A_1A_2;B_1B_2E_1E_2E_3}M_{E_1E_2E_3} = 0
\qquad {\cal M}_{6}^{AE_1;B_1B_2B_3E_2E_3}M_{E_1E_2E_3} = 0\]

\[{\cal M}_{6}^{E_1E_2;E_3B_1 \ldots B_4}M_{E_1E_2E_3} = 0 \]

\begin{eqnarray*}
{\cal M}_{6}^{E_1E_2;B_1 \ldots B_5}M^C \,\!_{E_1E_2}& = &0 \\
{\cal M}_{6}^{AE_1;B_1 \ldots B_4E_2}M^C \,\!_{E_1E_2}& = &
-\frac{3}{5} {\cal M}_{8}^{AC;B_1 \ldots B_4} \\
{\cal M}_{6}^{A_1A_2;B_1B_2B_3E_1E_2}M^C \,\!_{E_1E_2} &=&
\frac{1}{5} \left(4{\cal M}_{8}^{CA_1;A_2B_1B_2B_3}
-3 {\cal M}_{8}^{CA_1A_2;B_1B_2B_3} \right)
\end{eqnarray*}

\begin{eqnarray*}
{\cal{M}}_{6}^{B_1B_2;AB_3B_4B_5E}M^{CB_6} \,\!_E &=&
\frac{1}{20 \times 5!} \epsilon^{B_1 \ldots B_6E_1 \ldots E_4}
{\cal M}_{8}^{CA;} \,\!_{E_1 \ldots E_4} \\
{\cal{M}}_{6}^{AB_1;B_2 \ldots B_5E}M^{CB_6} \,\!_E &=&
- 2{\cal{M}}_{6}^{B_1B_2;AB_3B_4B_5E}M^{CB_6} \,\!_E
\end{eqnarray*}

\[{\cal{M}}_{6}^{AE;B_1 \ldots B_5}M^{C_1C_2} \,\!_E =
\frac{6}{5} \left(\frac{1}{5!} \epsilon^{B_1 \ldots B_5C_1E_1 \ldots E_4}
{\cal M}_{8}^{AC_2;} \,\!_{E_1 \ldots E_4}
- \eta^{B_1C_1} {\cal M}_{8}^{AC_2;B_2 \ldots B_5} \right) \]

\begin{eqnarray*}
\lefteqn{{\cal{M}}_{6}^{A_1A_2;B_1 \ldots B_4E}M^{C_1C_2} \,\!_E =
-\frac{3}{5 \times 5!} \epsilon^{B_1 \ldots B_4A_1C_1E_1 \ldots E_4}
{\cal M}_{8}^{A_2C_2;} \,\!_{E_1 \ldots E_4} } \\
& &+\frac{24}{25} \left[ \eta^{B_1C_1} {\cal M}_{8}^{C_2A_1;A_2B_2B_3B_4}
- \frac{1}{4}
\eta^{A_1C_1} {\cal M}_{8}^{A_2C_2;B_1 \ldots B_4}
- \frac{1}{3}
\eta^{A_1B_1} {\cal M}_{8}^{A_2C_1;C_2B_2B_3B_4}  \right] \\
& & -\frac{3}{10} \left(
3 \eta^{B_1C_1} {\cal M}_{8}^{C_2A_1A_2;B_2B_3B_4}
+\eta^{A_1B_1} {\cal M}_{8}^{A_2C_1C_2;B_2B_3B_4} \right)
\end{eqnarray*}

\[ {\cal{M}}_{6}^{C_1C_2;B_1 \ldots B_5}M^{C_3B_6B_7} =
 \frac{2}{7 \times 5!} \epsilon^{B_1 \ldots B_7E_1E_2E_3}
{\cal M}_{8}^{C_1C_2C_3;} \,\!_{E_1E_2E_3} \]

\begin{eqnarray}
\lefteqn{{\cal{M}}_{6}^{A_1A_2;B_1 \ldots B_5}M^{C_1C_2C_3} = }
\nonumber \\
&=& \frac{1}{32 \times 35} \bigl(
\epsilon^{B_1 \ldots B_5A_1A_2E_1E_2E_3}
{\cal M}_{8}^{C_1C_2C_3;} \,\!_{E_1E_2E_3}
+ 15 \epsilon^{B_1 \ldots B_5C_1C_2E_1E_2E_3}
{\cal M}_{8}^{A_1A_2C_3;} \,\!_{E_1E_2E_3}
\nonumber \\
& & - 12 \epsilon^{B_1 \ldots B_5A_1C_1E_1E_2E_3}
{\cal M}_{8}^{A_2C_2C_3;} \,\!_{E_1E_2E_3} \bigr)
\nonumber \\
& & -\frac{1}{16 \times 5} \biggl(
\frac{4}{5} \eta^{A_1C_1} \epsilon^{B_1 \ldots B_5C_2E_1 \ldots E_4}
{\cal M}_{8}^{A_2C_3;} \,\!_{E_1 \ldots E_4}
+ \eta^{B_1C_1} \epsilon^{B_2 \ldots B_5A_1C_2E_1 \ldots E_4}
{\cal M}_{8}^{A_2C_3;} \,\!_{E_1 \ldots E_4}
\nonumber \\
& & - \eta^{A_1B_1} \epsilon^{B_2 \ldots B_5C_1C_2E_1 \ldots E_4}
{\cal M}_{8}^{A_2C_3;} \,\!_{E_1 \ldots E_4} \biggr)
\nonumber \\
& & + \frac{6}{5} \bigl(
\eta^{B_1C_1} \eta^{B_2C_2} {\cal M}_{8}^{C_3A_1;A_2B_3B_4B_5}
+\frac{3}{4} \eta^{A_1C_1} \eta^{B_1C_2} {\cal M}_{8}^{C_3A_2;B_2 \ldots B_5}
\nonumber \\
& & +\eta^{B_1A_1} \eta^{B_2C_1} {\cal M}_{8}^{A_2C_2;C_3B_3B_4B_5} \bigr)
\nonumber \\
& &- \frac{3}{7} \biggl(
\frac{15}{4} \eta^{B_1C_1} \eta^{B_2C_2} {\cal M}_{8}^{C_3A_1A_2;B_3B_4B_5}
- 3\eta^{B_1A_1} \eta^{B_2C_1} {\cal M}_{8}^{A_2C_2C_3;B_3B_4B_5}
\nonumber \\
& & + \frac{1}{4}
\eta^{A_1B_1} \eta^{A_2B_2} {\cal M}_{8}^{B_3B_4B_5;C_1C_2C_3} \biggr)
\end{eqnarray}
\ve

\section{$\theta^{3}$-Fierz Identity and $\Gamma$-tracelessness.}
\setcounter{equation}{0}

\b

The basic Fierz identity does not need to have four $\theta$'s but only three.
Thus, (2.1) can be derived from

\be
\theta^{(\pm)} \bar\theta^{\pm)} {\cal O} \theta^{(\pm)} =
\frac{1}{96} \Pi^{(\pm)} \Gamma^{B_1B_2B_3} {\cal O} \theta^{(\pm)}
\bar\theta^{(\pm)} \Gamma_{B_1B_2B_3} \theta^{(\pm)}
\ee

\no An immediate consequence of (5.1) is

\be
\Gamma^{B_1B_2B3} \theta^{(\pm)}
\bar\theta^{(\pm)} \Gamma_{B_1B_2B_3} \theta^{(\pm)} = 0
\ee

\no and using (5.1) and (5.2) one easily obtains

\be
\Gamma^{B_1B_2} \theta^{(\pm)}
\bar\theta^{(\pm)} \Gamma_{B_1B_2A} \theta^{(\pm)} = 0
\ee

Then one can finally Fierz the general uncontracted product to obtain

\be
\theta^{(\pm)}
\bar\theta^{(\pm)} \Gamma_{A_1A_2A_3} \theta^{(\pm)} =
\frac{1}{2} \Gamma_{A_1}\Gamma^{B} \theta^{(\pm)}
\bar\theta^{(\pm)} \Gamma_{A_2A_3B} \theta^{(\pm)}
\ee

\no after using (5.1-5.3) and the properties of the Dirac algebra. Eq. (5.4)
gives us the decomposition of the product ${\cal M}^{A_{1} A_{2} A_{3}} \theta$
into irreducible pieces, and we see that the $\theta^{3}$ irreducible
spinor-tensor corresponding to $\left[\frac{3}{2} \frac{3}{2} \frac{1}{2}
\frac{1}{2} \frac{-1}{2} \right]$ is

\be
\Theta_{3}^{A_1A_2} = \Gamma_{E} M^{A_1A_2E} \theta
\ee

\no which is obviously traceless and by (5.3) also $\Gamma$-traceless. Thus,
(5.4) means

\be
M^{A_1A_2A_3} \theta = \frac{1}{2} \Gamma^{A_1} \Theta_{3}^{A_2A_3}
\ee

Of course, this decomposition can be obtained easily by detracing and
Young-projecting,

\be
M^{A_1A_2A_3} \theta = Traceless(M^{A_1A_2A_3} \theta)
+ a  \Gamma^{[A_1} \Gamma_EM^{A_2A_3]E} \theta
\ee

\no where ``Traceless" now means both $\eta$- and $\Gamma$-traceless and there
are no $\eta$ terms on the r.h.s. because the l.h.s. is trivially
$\eta$-traceless. But the Traceless term in (5.7) vanishes because there are no
irreducible objects with 3 tensor indices in the $\theta^{3}$ sector. The
constant $a$ is easily determined by contracting (5.7) with $\Gamma_{A_{1}}$,
to get $a = \frac{1}{2}$ and therefore reobtaining (5.6). The fermionic version
of the Young-projector mentioned in the previous paragraph is straightforward
enough, but it can become quite complicated for higher order decompositions. In
order to simplify things, the general way to proceed is as follows. First, we
figure out the irreducible objects by contracting as many indices as possible
in the product ${\cal M}^{A_{1} A_{2} A_{3}} \Theta_{n}$ so that the number of
remaining tensor indices are equal to the number of boxes of the corresponding
Young-pattern, and then we apply the Young-projector to the resulting object.
Next, we decompose the ${\cal M}_{n + 1} \theta$ products in terms of those
irreducible pieces instead of decomposing $M^{A_{1} A_{2} A_{3}} \Theta_{n}$
since the former is much easier than the latter in general. Finally, we may use
the results of the bosonic decompositions to obtain the decomposition of
$M^{A_{1} A_{2} A_{3}} \Theta_{n}$, since every fermionic irreducible object
$\Theta_{n}$ is expressed as some $\Gamma$-contraction of
${\cal M}_{n - 1} \theta$. The procedure will be illustrated in the first few
examples of the next section.
\ve

\section{ Irreducible Spinor-Tensors.}
\setcounter{equation}{0}

\b

Unlike in the bosonic case, this time we will proceed forward.

\vskip.25in
\no$\underline{\theta^{5}}$.

It is easy to obtain the anti-selfdual spinor-tensor corresponding to
$\left[\frac{3}{2} \frac{3}{2} \frac{3}{2} \frac{3}{2} \frac{-3}{2} \right]$:

\begin{eqnarray}
\Theta_{5}^{A_1 \ldots A_5} &=& M^{A_1A_2A_3} \Theta_{3}^{A_4A_5} =
M^{A_1A_2A_3}M^{A_4A_5C} \Gamma_C \theta =
\nonumber \\
&=& \Gamma_C {\cal M}_{4}^{C;A_1 \ldots A_5} \theta
\end{eqnarray}

\no Evidently it is traceless, but it is also $\Gamma$-traceless:

\begin{eqnarray}
\Gamma_D \Theta_{5}^{DA_1 \ldots A_4} &=&
\frac{1}{5} \Gamma_D \left(3 M^{DA_1A_2}M^{A_3A_4C}
+ 2 M^{A_1A_2A_3}M^{A_4DC} \right) \Gamma_C \theta
\nonumber \\
&=& \frac{3}{5} M^{DA_1A_2}M^{A_3A_4C} \Gamma_{DC} \theta = 0
\end{eqnarray}

\no where we have used (5.3) as well as (\ref{friend}). The anti-selfduality

\be
\Theta_{5}^{A_1 \ldots A_5} =
-\frac{1}{5!} \epsilon^{A_1 \ldots A_5 B_1 \ldots B_5}
\Theta_{5}\,\!_{B_1 \ldots B_5}
\ee

\no together with (6.2) imply the property

\be
\Gamma^{[B}\Theta_{5}^{A_1 \ldots A_5]} = 0
\ee

The second irreducible $\theta^{5}$ piece is:

\begin{eqnarray}
\Theta_{5}^{A;B_1B_2} &=& M^{B_1B_2} \,\!_E \Theta_{3}^{AE} =
M^{B_1B_2} \,\!_E M^{AED} \Gamma_D \theta
\nonumber \\
&=& {\cal M}_{4}^{DA;B_1B_2} \Gamma_D \theta
\end{eqnarray}

\no Usual tracelessness is also obvious here, while

\be
\Gamma_D \Theta_{5}^{D;B_1B_2} = 0
\ee

\no follows again from (5.3). The other $\Gamma$-trace also vanishes:

\begin{eqnarray}
\Gamma_D \Theta_{5}^{A;DB} &=&
\Gamma_D M^{DB} \,\!_E M^{AEF} \Gamma_F \theta
= M^{DB} \,\!_E M^{AEF} \Gamma_{DF} \theta
\nonumber \\
&=& M^{FD} \,\!_E M^{AEB} \Gamma_{DF} \theta = 0
\end{eqnarray}

\no where we used our old friend (\ref{friend}) and (5.3) once more. Lastly,
a property inherited from ${\cal M}_{4}^{A_{1} A_{2}; B_{1} B_{2}}$ is

\be
\Theta_{5}^{[A;B_1B_2]} = 0
\ee

Next we proceed to decompose products. By detracing one readily arrives at

\begin{eqnarray}
{\cal M}_{4}^{A_1A_2;B_1B_2} \theta &=&
\frac{1}{5} \left[
\Gamma^{A_1} \Theta_{5}^{A_2;B_1B_2} +
\Gamma^{B_1} \Theta_{5}^{B_2;A_1A_2} \right] \\
{\cal M}_{4}^{A;B_1 \ldots B_5} \theta &=&
\frac{1}{10} \left(\Gamma^A \Theta_{5}^{B_1 \ldots B_5}
+ \Gamma^{B_1B_2B_3} \Theta_5^{A;B_4B_5} \right)
\end{eqnarray}

\no With (6.9), (6.10) and (3.5) one can write the more general product

\begin{eqnarray}
M^{A_1A_2A_3} M^{B_1B_2B_3} \theta &=&
\frac{1}{2}
\Gamma^{A_1} \Theta_{5}^{A_2A_3B_1B_2B_3}
\nonumber \\
& &+ \frac{3}{20}
\left[\Gamma^{A_1}\Gamma^{B_1B_2} \Theta_{5}^{B_3;A_2A_3}
+ \Gamma^{B_1} \Gamma^{A_1A_2}\Theta_{5}^{A_3;B_2B_3} \right]
\end{eqnarray}

\no from which in turn we get

\begin{eqnarray}
M^{A_1A_2A_3} \Theta_{3}^{B_1B_2} &=&
\Theta_{5}^{A_1A_2A_3B_1B_2}
- \frac{3}{10} \Gamma^{A_1} \Gamma^{B_1} \Theta_{5}^{B_2;A_2A_3}
\nonumber \\
&+& \frac{3}{10} \Gamma^{A_1A_2} \Theta_{5}^{A_3;B_1B_2}
- \frac{6}{10} \eta^{A_1B_1} \Theta_{5}^{B_2;A_2A_3}
\end{eqnarray}

\no$\underline{\theta^{7}}$.

For the representation
$\left[\frac{5}{2} \frac{3}{2} \frac{3}{2} \frac{1}{2} \frac{-1}{2} \right]$
we need an object with 4 tensor indices, so consider

\begin{eqnarray}
\hat{\Theta}_{7}^{A_1A_2;B_1B_2} &=& M^{A_1A_2} \,\!_E \Theta_{5}^{E;B_1B_2} =
M^{A_1A_2} \,\!_D M^{B_1B_2} \,\!_E \Theta_{3}^{DE} =
\nonumber \\
&=&\Gamma_C M^{A_1A_2} \,\!_E {\cal M}_{4}^{CE;B_1B_2} \theta
= 3 \Gamma_C {\cal M}_{6}^{CA_1;A_2B_1B_2} \theta
\end{eqnarray}

\no This object is evidently antisymmetric in $A_{1}, A_{2}$ and in $B_{1},
B_{2}$, but it is also antisymmetric upon interchange of both sets of indices:

\be
\hat{\Theta}_{7}^{A_1A_2;B_1B_2} =
- \hat{\Theta}_{7}^{B_1B_2;A_1A_2}
\ee

\no Normal tracelessness is obvious and $\Gamma$-tracelessness follows from
that of $\Theta_{5}^{A; B_{1} B_{2}}$:

\be
\Gamma_E \hat{\Theta}_{7}^{EA;B_1B_2} =
\Gamma_E \hat{\Theta}_{7}^{B_1B_2;AE} =0
\ee

Also, from the definition we extract the properties

\begin{eqnarray}
\hat{\Theta}_{7}^{A[B;C]D} &=& \hat{\Theta}_{7}^{D[B;C]A}
\nonumber \\
\hat{\Theta}_{7}^{[A_1A_2;B_1B_2]} &=& 0
\end{eqnarray}

\no Clearly, this object must be irreducible; however, the corresponding Young
pattern symmetry is not manifest, so we define the new object

\be
\Theta_{7}^{B;A_1A_2A_3} =
\hat{\Theta}_{7}^{B[A_1;A_2A_3]}
= \Gamma_C {\cal M}_{6}^{CB;A_1A_2A_3} \theta
\ee

\no Eq. (6.16) implies

\be
\Theta_{7}^{[B;A_1A_2A_3]} = 0
\ee

\no
Again, these two spinor-tensors are equivalent and the inverse of (6.17) is

\be
\hat{\Theta}_{7}^{A_1A_2;B_1B_2} =
-3 \Theta_{7}^{[B_1;B_2]A_1A_2}
\ee

For the representation
$\left[\frac{7}{2} \frac{1}{2} \frac{1}{2} \frac{1}{2} \frac{1}{2} \right]$
we need an object with 3 tensor indices, so try

\begin{eqnarray}
\Theta_{7}^{ABC} &=& \Gamma_D M^{DA} \,\!_E \Theta_{5}^{C;BE} =
\Gamma_D M^{DAE} M^{B} \,\!_{FE} \Theta_{3}^{FC} =
\nonumber \\
&=&\Gamma_D {\cal M}_{4}^{DA;B} \,\!_F \, \Theta_{3}^{FC}
\end{eqnarray}

\no
{}From (6.20), (5.5), (4.80), (4.85) and the properties of ${\cal M}_{6}^{S_{1}
S_{2}; B_{1} B_{2} B_{3}}$ one can also obtain

\be
\Theta_{7}^{ABC} =
- \frac{3}{2} \Gamma_{D_1D_2} {\cal M}_{6}^{B(A;C)D_1D_2} \theta
\ee

\no which shows that $\Theta_{7}^{A B C}$ is symmetric in $A, C$. In order to
show that it is completely symmetric, we need to prove symmetry in $A, B$:

\begin{eqnarray}
\Theta_{7}^{[AB]C}
&=&-\frac{1}{2} \Gamma_D M^{ABE} M^{D} \,\!_{FE} \Theta_{3}^{FC}
\nonumber \\
&=& -\frac{1}{2} M^{AB} \,\!_E \Gamma_D \Theta_{5}^{C;DE} = 0
\end{eqnarray}

\no Thus:

\be
\Theta_{7}^{ABC} = \Theta_{7}^{BAC} = \Theta_{7}^{CBA} = \Theta_{7}^{ACB}
\ee

Next let us show that it vanishes upon contraction with $\Gamma_{A}$,

\begin{eqnarray}
\Gamma_C \Theta_{7}^{ABC} &=&
\Gamma_C \Gamma_D M^{DAE} M^{B} \,\!_{FE} \Theta_{3}^{FC}
\nonumber \\
&=&2 M_{C} \,\!^{AE} M^{B} \,\!_{FE} \Theta_{3}^{FC}
= - M^{BAE} M_{FCE} \Theta_{3}^{FC} = 0
\end{eqnarray}

\no as it is clear from (5.5) and (2.3).

Now we proceed to list the $\theta^{6} \times \theta$ decompositions. First,
by Young projection we get

\be
\Gamma_{E_1E_2} {\cal M}_{6}^{S_1S_2;CE_1E_2} \theta = - \frac{2}{3}
\Theta_{7}^{S_1S_2C}
\ee

\no which can also be obtained from (6.21) plus (6.23). For the remaining
${\cal M}_{6}^{S_{1} S_{2}; B_{1} B_{2} B_{3}} \theta$ products we have,
together with (6.17),

\begin{eqnarray}
\Gamma_{E_1E_2} {\cal M}_{6}^{SE_1;E_2B_1B_2} \theta &=& 0
\nonumber \\
\Gamma_{E} {\cal M}_{6}^{S_1S_2;C_1C_2E} \theta &=&
\frac{1}{2} \Theta_{7}^{S_1;S_2C_1C_2} +
\frac{1}{6} \Gamma^{C_1} \Theta_{7}^{C_2S_1S_2}
\nonumber \\
{\cal M}_{6}^{S_1S_2;B_1B_2B_3} \theta &=&
\frac{1}{7} \Gamma^{S_1} \Theta_{7}^{S_2;B_1B_2B_3}
+ \frac{3}{28} \Gamma^{B_1} \Theta_{7}^{S_1;S_2B_2B_3}
+ \frac{1}{28} \Gamma^{B_1B_2} \Theta_{7}^{B_3S_1S_2}
\end{eqnarray}

For ${\cal M}_{6}^{S_{1} S_{2}; B_{1}...B_{5}} \theta$ we have instead:

\begin{eqnarray*}
\Gamma_{E_1 \ldots E_4} {\cal M}_{6}^{A_1A_2;CE_1 \ldots E_4} \theta = 0&
&\Gamma_{E_1 \ldots E_4} {\cal M}_{6}^{AE_1;B_1B_2E_2E_3E_4} \theta = 0 \\
\Gamma_{E_1E_2E_3} {\cal M}_{6}^{AE_1;E_2E_3B_1B_2B_3} \theta =
- \frac{6}{5} \Theta_{7}^{A;B_1B_2B_3} &
&\Gamma_{E_1E_2E_3} {\cal M}_{6}^{A_1A_2;B_1B_2E_1E_2E_3} \theta =
- \frac{18}{5} \Theta_{7}^{A_1;A_2B_1B_2}
\end{eqnarray*}

\[ \Gamma_{E_1E_2} {\cal M}_{6}^{E_1E_2;B_1 \ldots B_5} \theta = 0\]

\begin{eqnarray*}
\Gamma_{E_1E_2} {\cal M}_{6}^{AE_1;E_2B_1 \ldots B_4} \theta =
- \frac{6}{5} \Gamma^{B_1} \Theta_{7}^{A;B_2B_3B_4} &
&\Gamma_{E_1E_2} {\cal M}_{6}^{A_1A_2;B_1B_2B_3E_1E_2} \theta =
- \frac{9}{5} \Gamma^{B_1} \Theta_{7}^{A_1;A_2B_2B_3}
\end{eqnarray*}

\begin{eqnarray*}
\Gamma_{E} {\cal M}_{6}^{AE;B_1 \ldots B_5} \theta &=&
\Gamma^{B_1B_2} \Theta_{7}^{A;B_3B_4B_5} \\
\Gamma_{E} {\cal M}_{6}^{A_1A_2;B_1 \ldots B_4E} \theta &=&
\frac{2}{35} \bigl[ \Gamma^{A_1}\Gamma^{B_1} \Theta_{7}^{A_2;B_2B_3B_4}
+ 12\Gamma^{B_1B_2} \Theta_{7}^{A_1;A_2B_3B_4} \\
& &+2 \eta^{A_1B_1} \Theta_{7}^{A_2;B_2B_3B_4} \bigr]
\end{eqnarray*}

\be
{\cal M}_{6}^{A_1A_2;B_1 \ldots B_5} \theta =
- \frac{1}{7} \Gamma^{A_1}\Gamma^{B_1B_2} \Theta_{7}^{A_2;B_3B_4B_5}
+ \frac{3}{14} \Gamma^{B_1B_2B_3} \Theta_{7}^{A_1;A_2B_4B_5}
\ee

\no$\underline{\theta^{9}}$.

In this sector, we have the same representations than in the previous
($\theta^{7}$) one. Inspired by (6.16), one defines

\begin{eqnarray}
\Theta_{9}^{ABC} &=& M^{A} \,\!_{DE} \hat{\Theta}_{7}^{BD;EC} =
\frac{3}{2} \Gamma_F M^{A} \,\!_{DE} {\cal M}_{6}^{F(B;C)DE} \theta
\nonumber \\
&=& - \Gamma_F {\cal M}_{8}^{FABC} \theta
\end{eqnarray}

Its tracelessness and total symmetry have become obvious in the last
equality in (6.28); hence this is the irreducible spinor-tensor corresponding
to $\left[\frac{7}{2} \frac{1}{2} \frac{1}{2} \frac{1}{2} \frac{-1}{2}
\right]$.
By projecting the product $M^{A_{1} A_{2}} \,\!_{D} \Theta_{7}^{S_{1} S_{2} D}$
one realizes that the other irreducible structure must be

\setcounter{equation}{0}
\newcounter{subequation}
\setcounter{subequation}{29}
\renewcommand{\theequation}
{\arabic{section}.\arabic{subequation}\alph{equation}}

\begin{eqnarray}
\Theta_{9}^{B;A_1A_2A_3} &=& M^{A_1A_2} \,\!_{D} \Theta_{7}^{A_3BD} \nonumber
\\
&=&- \frac{3}{2} \Gamma_{E_1E_2} M^{A_1A_2} \,\!_{D}
{\cal M}_{6}^{A_3D;BE_1E_2} \theta
=\frac{3}{2} \Gamma_{E_1E_2} {\cal M}_{8}^{BE_1E_2;A_1A_2A_3} \theta.
\end{eqnarray}

Exploiting the symmetry of $\Theta_{7}^{ABC}$ we can interchange the roles
of $A_3$ and $B$ in (6.29a) and using the properties of
${\cal M}_{8}^{S_1S_2;D_1 \ldots D_4}$ as well as the last equality in (6.29a),
one can equally derive

\be
\Theta_{9}^{B;A_1A_2A_3} =
2 \Gamma_{EF} {\cal M}_{8}^{BE;FA_1A_2A_3} \theta.
\ee

The ordinary trace vanishes manifestly as does the first $\Gamma$-trace:

\setcounter{equation}{29}
\renewcommand{\theequation}{\arabic{section}.\arabic{equation}}

\be
\Gamma_E \Theta_{9}^{E;A_1A_2A_3} = 0
\ee

\no The other one also vanishes:

\begin{eqnarray}
\Gamma_E \Theta_{9}^{B;EA_1A_2} &=&
\frac{2}{3} \Gamma_E M^{EA_1} \,\!_D \Theta_{7}^{A_2BD} =
\frac{2}{3} \Gamma_E M^{EA_1} \,\!_D \Gamma_F M^{FA_2} \,\!_C
\Theta_{5}^{B;DC} = \nonumber \\
&=&\frac{2}{3} M^{EA_1} \,\!_D M_E \,\!^{A_2} \,\!_C \Theta_{5}^{B;DC} =
- \frac{1}{3} M^{EA_2A_1} M_{EDC} \Theta_{5}^{B;DC} = 0
\end{eqnarray}

\no as implied by (6.5) and (2.3). The remaining property inherited from
(4.75) is

\be
\Theta_{9}^{[B;A_1A_2A_3]} = 0
\ee

Turning to the $\theta^{8} \times \theta$ decompositions, the first one is
trivially inferred from (6.28)

\be
{\cal M}_{8}^{S_1S_2S_3S_4} \theta =
- \frac{1}{4} \Gamma^{S_1} \Theta_{9}^{S_2S_3S_4}.
\ee

\no
{}From (6.29a) one successively derives the set:

\begin{eqnarray}
\Gamma_E {\cal M}_{8}^{A_1A_2A_3;B_1B_2E} \theta &=&
- \frac{2}{15} \Gamma^{B_1} \Theta_{9}^{B_2;A_1A_2A_3}
+ \frac{2}{10} \Gamma^{A_1} \Theta_{9}^{B_1;B_2A_2A_3}
\nonumber \\
{\cal M}_{8}^{A_1A_2A_3;B_1B_2B_3} \theta &=&
- \frac{1}{45} \left( \Gamma^{A_1A_2} \Theta_{9}^{A_3;B_1B_2B_3}
+ \Gamma^{B_1B_2} \Theta_{9}^{B_3;A_1A_2A_3} \right)
\nonumber \\
&+& \frac{1}{15} \Gamma^{A_1B_1} \Theta_{9}^{A_2;A_3B_2B_3}
\end{eqnarray}

\no while from (6.29b) instead, the set

\begin{eqnarray}
\Gamma_E {\cal M}_{8}^{EA;B_1 \ldots B_4} \theta &=&
- \frac{1}{2} \Gamma^{B_1} \Theta_{9}^{A;B_2B_3B_4}
\nonumber \\
{\cal M}_{8}^{AC;B_1 \ldots B_4} \theta &=&
- \frac{2}{63} \left( \Gamma^{A} \Gamma^{B_1} \Theta_{9}^{C;B_2B_3B_4}
+ \Gamma^{C} \Gamma^{B_1} \Theta_{9}^{A;B_2B_3B_4} \right)
\nonumber \\
&+& \frac{1}{126} \left( \eta^{AB_1} \Theta_{9}^{C;B_2B_3B_4}
+ \eta^{CB_1} \Theta_{9}^{A;B_2B_3B_4} \right)
\nonumber \\
&+& \frac{1}{42} \Gamma^{B_1B_2}\left(  \Theta_{9}^{A;CB_3B_4}
+ \Theta_{9}^{C;AB_3B_4} \right)
+ \frac{1}{42} \Gamma^{B_1B_2B_3} \Theta_{9}^{B_4AC}
\end{eqnarray}

\no$\underline{\theta^{11}}$.

For the representation $\left[ \frac{5}{2} \frac{3}{2} \frac{1}{2}
\frac{1}{2} \frac{1}{2}\right]$
we first construct the object with 3 indices by contracting
$M^{A_1A_2A_3}$ with $\Theta_{9}^{ABC}$. We define

\begin{eqnarray}
\hat{\Theta}_{11}^{A;BC} &=& \Gamma_{D} M^{DAE}\Theta_9^{BC}\,\!_E
= \Gamma_D{\cal{M}}_8^{DBCE}\Theta_3\,\!^A\,\!_E \nonumber \\
&=& \Gamma_{E_1E_2}\hat{\cal{M}}_{10}^{BCE_1;E_2A}\theta
= \frac{3}{2} \Gamma_{E_1E_2} {\cal{M}}_{10}^{BC;AE_1E_2}\theta
\end{eqnarray}

\no Then, we see that the tracelessness of $\hat{\Theta}_{11}^{A;BC}$ is
trivially satisfied and the $\Gamma$-tracelessness is also immediate from
(6.36):

\[\Gamma_A\hat{\Theta}_{11}^{A;BC} =
\Gamma_A\Gamma_D{\cal{M}}_{8}^{DBCE}\Theta_3\,\!^A\,\!_E
= 2\eta_{AD}{\cal{M}}_{8}^{DBCE}\Theta_3\,\!^A\,\!_E = 0\]

\be
\Gamma_B\hat{\Theta}_{11}^{A;BC} =
\Gamma_B\Gamma_D
M^{DAE}\Theta_9^{BC}\,\!_E = 2 \eta_{BD}M^{DAE}\Theta_9^{BC}\,\!_E = 0
\ee

\no So $\hat{\Theta}_{11}^{A;BC}$ is irreducible, and a useful property of
$\hat{\Theta}_{11}^{A;BC}$ can be inferred from the group theory: i.e., we
must have

\be
\hat{\Theta}_{11}^{(A;BC)} = 0,
\ee

\no which reflects the fact that we can not have an irreducible object
with totally symmetrized 3 indices in $\theta^{11}$-sector (see Table1).
In fact, (6.38) can be readily verified from the definition (6.36):

\begin{eqnarray*}
\hat{\Theta}_{11}^{(A;BC)} &=&
- \Gamma_{E_1E_2}\hat{\cal{M}}_{10}^{E_1(BC;A)E_2}\theta
= \frac{1}{3}\Gamma_{E_1E_2}\hat{\cal{M}}_{10}^{BCA;E_1E_2}\theta \\
&=& \frac{1}{3} {\cal{M}}_{8}^{BCAF}\Gamma_{E_1E_2} M^{E_1E_2}\,\!_F\theta
= 0
\end{eqnarray*}

Even though $\hat{\Theta}_{11}^{A;BC}$ is irreducible, its Young symmetry
is not manifest, so we need to define a new object for $\left[\frac{5}{2}
\frac{3}{2} \frac{1}{2} \frac{1}{2} \frac{1}{2} \right]$:

\be
\Theta_{11}^{B;CD} = \hat{\Theta}_{11}^{[C;D]B}
= \frac{3}{2} \Gamma_{E_1E_2}{\cal{M}}_{10}^{BE_1;E_2CD}\theta.
\ee

\no Then, it is obvious from the definition (6.39) and (4.48) that
$\Theta_{11}^{B;A_1A_2}$ satisfies

\be
\Theta_{11}^{[B;A_1A_2]} = 0,
\ee

\no and the inverse of (6.39) is

\be
\hat{\Theta}_{11}^{A;S_1S_2} = -\frac{4}{3} \Theta_{11}^{S_1;S_2A}.
\ee

Turning to the representation $\left[\frac{3}{2} \frac{3}{2} \frac{3}{2}
\frac{3}{2} \frac{3}{2} \right]$, we need an object with 5 totally
antisymmetrized tensor indices. Naturally, we define

\be
\Theta_{11}^{A_1 \ldots A_5} = M_{E}\,\!^{A_1A_2}\Theta_9^{E;A_3A_4A_5}
= -\Gamma_{E_1E_2} {\cal{M}}_{10}^{E_1E_2;A_1 \ldots A_5} \theta.
\ee

\no Again, the tracelessness is trivial, but for the $\Gamma$-tracelessness
we need a little work:

\begin{eqnarray}
\Gamma_{A_1}\Theta_{11}^{A_1 \ldots A_5} &=&
\Gamma_{A_1} M^{E[A_1A_2} \Theta_{9}\,\!_{E}\,\!^{;A_3A_4A_5]}
= \frac{2}{5} \Gamma_{D} M^{DA_2}\,\!_{E} \Theta_9^{E;A_3A_4A_5}\nonumber\\
&=& \frac{2}{5} M^{A_2A_3}\,\!_F \Gamma_D M^{DA_4}\,\!_E \Theta_{7}^{A_5FE}
= \frac{3}{5} M^{A_2A_3}\,\!_F \Gamma_D \Theta_{9}^{F;DA_4A_5} = 0.
\end{eqnarray}

\no The irreducible object $\Theta_{11}^{A_1 \ldots A_5}$ satisfies
similar properties to those of $\Theta_5^{A_1 \ldots A_5}$. First, it is
self-dual:

\be
\Theta_{11}^{A_1 \ldots A_5} = {1\over 5!} \epsilon^{A_1 \ldots A_5B_1
\ldots B_5} \Theta_{11}\,\!_{B_1 \ldots B_5}
\ee

\no and it satisfies

\be
\Gamma^{[B} \Theta_{11}^{A_1 \ldots A_5]} = 0.
\ee

\no While the self-duality (6.44) is obvious from (4.60) and (6.42),
eq. (6.45) may be obtained from (6.43) and (6.44) similarly to the
case of $\Theta_5^{A_1 \ldots A_5}$. In fact, the property (6.45) as well
as (6.4) may be also justified by the fact that:
(1)$\Gamma^{[A_1}\Theta_{11}^{A_2 \ldots A_6]}$ and
$\Gamma^{[A_1}\Theta_{5}^{A_2 \ldots A_6]}$ are irreducible and, (2)we can
not have an irreducible object with 6 fully antisymmetrized indices in the
$\theta^{11}$- and $\theta^5$-sectors. $\Gamma^{[A_1}\Theta_{11}^{A_2
\ldots A_6]}$ is indeed irreducible because it is both $\eta$- and
$\Gamma$-traceless:

\be
\Gamma_{A_1}\Gamma^{[A_1} \Theta_{11}^{A_2 \ldots A_6]} = 0,
\ee

\no as can be seen by expanding the bracket.

Now let us list the $\theta^{10} \times \theta$ decompositions. For
${\cal{M}}_{10}^{S_1S_2;A_1A_2A_3}\theta$ products we first have (6.36),
(6.39) and

\be
\Gamma_{E_1E_2} {\cal{M}}_{10}^{S_1S_2;AE_1E_2}\theta =
-\frac{8}{9} \Theta_{11}^{S_1;S_2A}.
\ee

\no Then from these two we successively obtain the remaining
decompositions:

\[\Gamma_{E}{\cal{M}}_{10}^{EA;B_1B_2B_3}\theta =
-\frac{1}{3} \Gamma^{B_1} \Theta_{11}^{A;B_2B_3}\]

\[\Gamma_E {\cal{M}}_{10}^{S_1S_2;A_1A_2E} \theta =
\frac{4}{63}\biggl(\Gamma^{S_1}\Theta_{11}^{S_2;A_1A_2}
+ 3\Gamma^{A_1}\Theta_{11}^{S_1;S_2A_2}\biggr)\]

\be
{\cal{M}}_{10}^{S_1S_2;A_1A_2A_3}\theta =
\frac{1}{210}\biggl(-9\Gamma^{S_1}\Gamma^{A_1}\Theta_{11}^{S_2;A_2A_3}
+ 4\eta^{S_1A_1}\Theta_{11}^{S_2;A_2A_3}
+ 6\Gamma^{A_1A_2}\Theta_{11}^{S_1;S_2A_3}\biggr).
\ee

\no On the other hand, for ${\cal{M}}_{10}^{A_1A_2;B_1 \ldots B_5}\theta$
we have (6.42) and

\[\Gamma_{E_1 \ldots E_4} {\cal{M}}_{10}^{A_1A_2;BE_1 \ldots E_4}\theta
= -\frac{8}{5}\Theta_{11}^{B;A_1A_2}\]

\[\Gamma_{E_1 \ldots E_4} {\cal{M}}_{10}^{BE_1;E_2E_3E_4A_1A_2}\theta
= -\frac{2}{5}\Theta_{11}^{B;A_1A_2}\]

\[\Gamma_{E_1E_2E_3}{\cal{M}}_{10}^{A_1A_2;B_1B_2E_1E_2E_3}\theta
= \frac{2}{5}\Gamma^{B_1}\Theta_{11}^{B_2;A_1A_2}\]

\[\Gamma_{E_1E_2E_3}{\cal{M}}_{10}^{AE_1;E_2E_3B_1B_2B_3}\theta
= \frac{1}{5}\Gamma^{B_1}\Theta_{11}^{A;B_2B_3}\]

\[\Gamma_{E_1E_2E_3}{\cal{M}}_{10}^{E_1E_2;E_3A_1 \ldots A_4}\theta = 0\]

\[\Gamma_{E_1E_2}{\cal{M}}_{10}^{AE_1;E_2B_1 \ldots B_4}\theta
= \frac{1}{5}\Theta_{11}^{AB_1 \ldots B_4}
+ \frac{4}{50}\Gamma^{B_1B_2}\Theta_{11}^{A;B_3B_4}\]

\begin{eqnarray*}
\Gamma_{E_1E_2}{\cal{M}}_{10}^{A_1A_2;B_1B_2B_3E_1E_2}\theta &=&
-\frac{1}{10}\Theta_{11}^{A_1A_2B_1B_2B_3}\\
&+& \frac{1}{200}\biggl(17\Gamma^{B_1B_2}\Theta_{11}^{B_3;A_1A_2}
-\Gamma^{A_1}\Gamma^{B_1}\Theta_{11}^{A_2;B_2B_3}
-4\eta^{A_1B_1}\Theta_{11}^{A_2;B_2B_3}\biggr)
\end{eqnarray*}

\[\Gamma_E{\cal{M}}_{10}^{EA;B_1 \ldots B_5}\theta =
\frac{1}{10}\Gamma^A\Theta_{11}^{B_1 \ldots B_5}
+\frac{1}{30}\Gamma^{B_1B_2B_3}\Theta_{11}^{A;B_4B_5}\]

\begin{eqnarray*}
\Gamma_E{\cal{M}}_{10}^{A_1A_2;B_1 \ldots B_4E}\theta &=&
-\frac{1}{20}\Gamma^{A_1}\Theta_{11}^{A_2B_1 \ldots B_4}\\
&-&\frac{1}{600}\biggl(3\Gamma^{A_1}\Gamma^{B_1B_2}\Theta_{11}^{A_2;B_3B_4}
+11\Gamma^{B_1B_2B_3}\Theta_{11}^{B_4;A_1A_2}
+6\eta^{A_1B_1}\Gamma^{B_2}\Theta_{11}^{A_2;B_3B_4}\biggr)
\end{eqnarray*}

\begin{eqnarray}
{\cal{M}}_{10}^{A_1A_2;B_1 \ldots B_5}\theta &=&
\frac{1}{88}\biggl(\Gamma^{A_1A_2}\Theta_{11}^{B_1 \ldots B_5}
-2\eta^{A_1B_1}\Theta_{11}^{B_2 \ldots B_5A_2}\biggr) \nonumber \\
&+&\frac{1}{240}\biggl(\Gamma^{A_1}\Gamma^{B_1B_2B_3}
\Theta_{11}^{A_2;B_4B_5}
-\Gamma^{B_1 \ldots B_4}\Theta_{11}^{B_5;A_1A_2}\biggr)
\end{eqnarray}

\no$\underline{\theta^{13}}.$

The only representation we have in this sector is just $\left[\frac{3}{2}
\frac{3}{2} \frac{1}{2} \frac{1}{2} \frac{1}{2} \right]$ like in the
$\theta^3$-sector and this means that we need an object with 2
antisymmetric tensor indices again. Let us define

\be
\Theta_{13}^{AB} = M^{A}\,\!_{E_1E_2} \Theta_{11}^{B;E_1E_2}.
\ee

Then the antisymmetry property of $\Theta_{13}^{AB}$ is automatically
insured as soon as we obtain the following identity. That is, if we use
(6.39), (4.48) and the first equation of (4.49), eq. (6.50) becomes

\begin{eqnarray}
\Theta_{13}^{AB} &=&
\frac{3}{2} \Gamma_{D_1D_2} {\cal{M}}_{10}^{BD_1;D_2E_1E_2}
M^{A}\,\!_{E_1E_2}\theta \nonumber \\
&=& -\frac{3}{2} \Gamma_{D_1D_2} {\cal{M}}_{10}^{BE_1;E_2D_1D_2}
M^{A}\,\!_{E_1E_2}\theta \nonumber \\
&=& -\Gamma_{D_1D_2} {\cal{M}}_{12}^{D_1D_2;AB}\theta
\end{eqnarray}

\no Further, the other expression for $\Theta_{13}^{AB}$ is also
immediately obtained from (6.51) if we use the first equation of (4.61),
and (6.42):

\be
\Theta_{13}^{AB} =
-\frac{5}{2}\Gamma_{D_1D_2}{\cal{M}}_{10}^{D_1D_2;ABE_1E_2E_3}
M_{E_1E_2E_3}\theta
= \frac{5}{2} M_{E_1E_2E_3}\Theta_{11}^{E_1E_2E_3AB}.
\ee

On the other hand, the normal tracelessness of this antisymmetric
spinor-tensor is trivial and

\be
\Gamma_D\Theta_{13}\,\!^{DA} = 0
\ee

\no is also obvious from the last equality in (6.52). So
$\Theta_{13}^{A_1A_2}$ is the irreducible object corresponding to the
representation $\left[\frac{3}{2} \frac{3}{2} \frac{1}{2} \frac{1}{2}
\frac{1}{2} \right]$.
Now, for the $\theta^{12}\times\theta$ decompositions we have

\[\Gamma_E {\cal{M}}_{12}^{EA;B_1B_2}\theta =
\frac{1}{12} \biggl(\Gamma^A\Theta_{13}^{B_1B_2}
-\Gamma^{B_1}\Theta_{13}^{B_2A}\biggr)\]

\[ \hat{\cal M}_{12}^{S_1S_2;X_1X_2}\theta=
\frac{1}{33} \Gamma^{S_1X_1}\Theta_{13}^{S_2X_2} \]

\be
{\cal{M}}_{12}^{A_1A_2;B_1B_2}\theta=
\frac{1}{132} \biggl(\Gamma^{A_1A_2}\Theta_{13}^{B_1B_2}
+\Gamma^{B_1B_2}\Theta_{13}^{A_1A_2}
+2\Gamma^{A_1B_1}\Theta_{13}^{A_2B_2}\biggr)
\ee

\no and

\[\Gamma_{E_1 \ldots E_5} {\cal{M}}_{12}^{A;E_1 \ldots E_5}\theta = 0\]

\[\Gamma_{E_1 \ldots E_4} {\cal{M}}_{12}^{A;BE_1 \ldots E_4}\theta =
-\frac{4}{5} \Theta_{13}^{AB}\]

\[\Gamma_{E_1 \ldots E_4} {\cal{M}}_{12}^{E_1;E_2E_3E_4A_1A_2}\theta =
\frac{2}{5} \Theta_{13}^{A_1A_2}\]

\[\Gamma_{E_1E_2E_3} {\cal{M}}_{12}^{B;A_1A_2E_1E_2E_3}\theta =
-\frac{3}{15} \Gamma^{A_1} \Theta_{13}^{A_2B}\]

\[\Gamma_{E_1E_2E_3}{\cal{M}}_{12}^{E_1;E_2E_3A_1A_2A_3}\theta=
-\frac{1}{5}\Gamma^{A_1}\Theta_{13}^{A_2A_3}\]

\[\Gamma_{E_1E_2} {\cal{M}}_{12}^{B;A_1A_2A_3E_1E_2}\theta =
-\frac{1}{450} \biggl(19\Gamma^{A_1A_2}\Theta_{13}^{A_3B}
+\Gamma^{B}\Gamma^{A_1}\Theta_{13}^{A_2A_3}
+4\eta^{BA_1}\Theta_{13}^{A_2A_3}\biggr)\]

\[\Gamma_{E_1E_2}{\cal{M}}_{12}^{E_1;E_2A_1 \ldots A_4}\theta =
-\frac{2}{25}\Gamma^{A_1A_2}\Theta_{13}^{A_3A_4}\]

\[\Gamma_E{\cal{M}}_{12}^{E;A_1 \ldots A_5}\theta =
\frac{1}{30}\Gamma^{A_1A_2A_3}\Theta_{13}^{A_4A_5}\]

\begin{eqnarray*}
\Gamma_E{\cal{M}}_{12}^{B;A_1 \ldots A_4E}\theta=
\frac{1}{450}\biggl(4\Gamma^{A_1A_2A_3}\Theta_{13}^{A_4B}
-\Gamma^B\Gamma^{A_1A_2}\Theta_{13}^{A_3A_4}
-2\eta^{BA_1}\Gamma^{A_2}\Theta_{13}^{A_3A_4}\biggr)
\end{eqnarray*}

\be
{\cal{M}}_{12}^{B;A_1 \ldots A_5}\theta=
\frac{1}{540} \biggl(\Gamma^B\Gamma^{A_1A_2A_3}\Theta_{13}^{A_4A_5}
+\Gamma^{A_1 \ldots A_4}\Theta_{13}^{A_5B}\biggr)
\ee

\no$\underline{\theta^{15}}$.

Finally, for $\theta^{15}$-sector we have again only one representation,
which is $\left[\frac{1}{2} \frac{1}{2} \frac{1}{2} \frac{1}{2} \frac{-1}{2}
\right]$ and the corresponding irreducible object is a spinor with no
tensor indices just like $\theta$, but with
opposite chirality in this case. So the only possible candidate for
$\Theta_{15}$ is:

\be
\Theta_{15} \equiv \Theta = \Gamma_{D} M^{DE_1E_2}\Theta_{13}\,\!_{E_1E_2}=
\Gamma_{E_1E_2E_3}{\cal{M}}^{E_1E_2E_3}\theta.
\ee

\no For the decompositions we have

\begin{eqnarray}
\Gamma_{E_1E_2}{\cal{M}}^{E_1E_2A}\theta &=& \frac{1}{10}\Gamma^A\Theta
\nonumber \\
\Gamma_E{\cal{M}}^{EA_1A_2}\theta &=& -\frac{1}{90}\Gamma^{A_1A_2}\Theta
\nonumber \\
{\cal{M}}^{A_1A_2A_3}\theta &=& -\frac{1}{720}\Gamma^{A_1A_2A_3}\Theta
\end{eqnarray}

\ve
\section{Products of $M^{A_1A_2A_3}$ with Spinor-Tensors}
\setcounter{equation}{0}

In this section we list the products of $M^{A_1A_2A_3}$ with all the
$\Theta_{n}$ of section VI, since they are another necessary ingredient
in the development of the tensor calculus. Other more esoteric product
identities are given in Appendix B.

\vskip.25in
\no$\underline{\theta^{5}}$

\begin{eqnarray}
M^{A_1A_2A_3}\Theta_3^{B_1B_2} &=& \Theta_5^{A_1A_2A_3B_1B_2} \nonumber \\
&-& \frac{3}{10} \Gamma^{A_1}\Gamma^{B_1}\Theta_5^{B_2;A_2A_3}
+ \frac{3}{10} \Gamma^{A_1A_2}\Theta_5^{A_3;B_1B_2}
- \frac{6}{10} \eta^{A_1B_1}\Theta_5^{B_2;A_2A_3}
\end{eqnarray}

\no$\underline{\theta^{7}}$

\begin{eqnarray}
M^{A_1A_2A_3}\Theta_5^{B_1 \ldots B_5} &=&
-\frac{1}{14} \Gamma^{B_1 \ldots B_4}\Theta_7^{B_5;A_1A_2A_3}
+ \frac{15}{14} \biggl(\Gamma^{A_1}\Gamma^{B_1B_2B_3}
- 2\eta^{A_1B_1}\Gamma^{B_2B_3}\biggr)\Theta_7^{B_4;B_5A_2A_3}\nonumber \\
&-& \frac{5}{7} \biggl(\Gamma^{A_1A_2}\Gamma^{B_1B_2}
+ 2\eta^{A_1B_1}\Gamma^{A_2}\Gamma^{B_2}
- 2\eta^{A_1B_1}\eta^{A_2B_2}\biggr)\Theta_7^{A_3;B_3B_4B_5}
\end{eqnarray}

\begin{eqnarray}
M^{A_1A_2A_3}\Theta_5^{C;B_1B_2} &=&
\frac{5}{14} \biggl(-\Gamma^C\Gamma^{A_1}
+ 4\eta^{CA_1}\biggr)\Theta_7^{A_2;A_3B_1B_2}
+ \frac{1}{21} \biggl(\Gamma^C\Gamma^{B_1}
- 4\eta^{CB_1}\biggr)\Theta_7^{B_2;A_1A_2A_3} \nonumber \\
&-& \frac{5}{56} \biggl(\Gamma^{A_1}\Gamma^{B_1}
- 10\eta^{A_1B_1}\biggr)\Theta_7^{B_2;CA_2A_3}
- \frac{5}{56} \biggl(5\Gamma^{A_1}\Gamma^{B_1}
- 2\eta^{A_1B_1}\biggr)\Theta_7^{C;B_2A_2A_3} \nonumber \\
&+& \frac{5}{28} \Gamma^{A_1A_2}\Theta_7^{A_3;CB_1B_2}
- \frac{15}{28} \Gamma^{A_1A_2}\Theta_7^{C;A_3B_1B_2}
- \frac{1}{21} \Gamma^{B_1B_2}\Theta^{C;A_1A_2A_3} \nonumber \\
&+& \frac{1}{28} \biggl(\Gamma^{A_1A_2}\Gamma^{B_1}
- 4\eta^{A_1B_1}\Gamma^{A_2}\biggr)\Theta_7^{A_3B_2C}
\end{eqnarray}

\no$\underline{\theta^{9}}$

\begin{eqnarray}
\lefteqn{M^{A_1A_2A_3}\Theta_7^{B;C_1C_2C_3}
= -\frac{1}{60} \Gamma^{A_1A_2A_3}\Theta_9^{B;C_1C_2C_3}
+ \frac{1}{140} \Gamma^{C_1C_2C_3}\Theta_9^{B;A_1A_2A_3}} \nonumber \\
& &+ \frac{1}{140} \biggl(\Gamma^B\Gamma^{C_1C_2}
+ \frac{2}{3} \eta^{BC_1}\Gamma^{C_2}\biggr)\Theta_9^{C_3;A_1A_2A_3}
+ \frac{1}{10} \eta^{BA_1}\Gamma^{A_2}\Theta_9^{A_3;C_1C_2C_3} \nonumber \\
& &- \frac{9}{280} \biggl(\Gamma^{A_1}\Gamma^{C_1C_2}
+ \frac{2}{3} \eta^{A_1C_1}\Gamma^{C_2}\biggr)\Theta_9^{C_3;BA_2A_3}
- \frac{1}{280}\biggl(23 \Gamma^{A_1} \Gamma^{C_1C_2}
- 22 \eta^{A_1C_1}\Gamma^{C_2}\biggr)\Theta_9^{B;C_3A_2A_3} \nonumber \\
& &+ \frac{3}{20} \Gamma^{A_1A_2}\Gamma^{C_1}
\Theta_9^{B;A_3C_2C_3}
- \frac{1}{20} \biggl(\Gamma^{C_1}\Gamma^{A_1A_2}
- 6\eta^{A_1C_1}\Gamma^{A_2}\biggr)\Theta_9^{A_3;BC_2C_3} \nonumber \\
& &+ \frac{1}{20} \biggl(\Gamma^B\Gamma^{A_1}\Gamma^{C_1}
+ \eta^{A_1C_1}\Gamma^{B} + \eta^{BA_1}\Gamma^{C_1}
- \eta^{BC_1}\Gamma^{A_1}\biggr)\Theta_9^{A_2;A_3C_2C_3} \nonumber \\
& &- \frac{1}{140} \biggl(\Gamma^{A_1A_2}\Gamma^{C_1C_2}
- 6\eta^{A_1C_1}\Gamma^{A_2}\Gamma^{C_2}
- 22\eta^{A_1C_1}\eta^{A_2C_2}\biggr)\Theta_9^{A_3C_3B}
\end{eqnarray}

\begin{eqnarray}
M^{A_1A_2A_3}\Theta_7^{S_1S_2S_3} &=&
- \frac{1}{16} \Gamma^{A_1A_2A_3}\Theta_9^{S_1S_2S_3}
+ \frac{3}{112} \biggl(5\Gamma^{A_1A_2}\Gamma^{S_1}
- 8\eta^{S_1A_1}\Gamma^{A_2}\biggr)\Theta_9^{A_3S_2S_3} \nonumber \\
&-& \frac{1}{7} \eta^{S_1S_2}\Theta_9^{S_3;A_1A_2A_3}
- \frac{3}{28} \biggl(\Gamma^{S_1}\Gamma^{A_1}
- 14\eta^{S_1A_1}\biggr)\Theta_9^{S_2;S_3A_2A_3}
\end{eqnarray}

\no$\underline{\theta^{11}}$

\begin{eqnarray}
M^{A_1A_2A_3}\Theta_9^{S_1S_2S_3} &=&
\frac{2}{70} \biggl(\Gamma^{A_1A_2}\Gamma^{S_1}
- 10\eta^{S_1A_1}\Gamma^{A_2}\biggr)\Theta_{11}^{S_2;S_3A_3} \nonumber \\
&-& \frac{3}{70} \biggl(\eta^{S_1S_2}\Gamma^{A_1}
- 2\eta^{S_1A_1}\Gamma^{S_2}\biggr)\Theta_{11}^{S_3;A_2A_3}
\end{eqnarray}

\begin{eqnarray}
\lefteqn{M^{A_1A_2A_3}\Theta_9^{C;B_1B_2B_3} =
\frac{26}{330} \Gamma^{A_1A_2}\Theta_{11}^{A_3B_1B_2B_3C}
+ \frac{31}{330} \Gamma^{B_1B_2}\Theta_{11}^{B_3A_1A_2A_3C} }\nonumber \\
& &+ \frac{1}{330} \biggl(21\Gamma^C\Gamma^{A_1}
+ 109\eta^{CA_1}\biggr)\Theta_{11}^{A_2A_3B_1B_2B_3}
- \frac{91}{330} \eta^{CB_1}\Theta_{11}^{B_2B_3A_1A_2A_3}
+ \frac{213}{330} \eta^{A_1B_1}\Theta_{11}^{A_2A_3B_2B_3C} \nonumber \\
& &+ \frac{1}{1680} \biggl(11\Gamma^{B_1B_2B_3}\Gamma^{A_1}
- 50\eta^{A_1B_1}\Gamma^{B_2B_3}\biggr)\Theta_{11}^{C;A_2A_3}
- \frac{1}{40} \Gamma^{A_1A_2A_3}\Gamma^{B_1}\Theta_{11}^{C;B_2B_3} \nonumber
\\
& &+ \frac{1}{120} \biggl(2\Gamma^C\Gamma^{A_1A_2}\Gamma^{B_1}
- 9\eta^{CA_1}\Gamma^{A_2}\Gamma^{B_1} - \eta^{CB_1}\Gamma^{A_1A_2}
- \eta^{B_1A_1}\Gamma^C\Gamma^{A_2}
- 18\eta^{CA_1}\eta^{A_2B_1}\biggr)\Theta_{11}^{A_3;B_2B_3} \nonumber \\
& &+ \frac{1}{1680} \biggl(- 11\Gamma^{A_1}\Gamma^C\Gamma^{B_1B_2}
- 30\eta^{CB_1}\Gamma^{B_2}\Gamma^{A_1}
+ 24\eta^{A_1B_1}\Gamma^C\Gamma^{B_2}
+ 140\eta^{CB_1}\eta^{B_2A_1}\biggr)\Theta_{11}^{B_3;A_2A_3} \nonumber \\
& &- \frac{1}{840} \biggl(\Gamma^{B_1B_2}\Gamma^{A_1A_2}
+ 42\eta^{A_1B_1}\Gamma^{A_2}\Gamma^{B_2}
+ 168\eta^{A_1B_1}\eta^{A_2B_2}\biggr)\Theta_{11}^{B_3;A_3C} \nonumber \\
& &- \frac{1}{840}\biggl(29\Gamma^{A_1A_2}\Gamma^{B_1B_2}
+ 22\eta^{A_1B_1}\Gamma^{A_2}\Gamma^{B_2}
- 64\eta^{A_1B_1}\eta^{A_2B_2}\biggr)\Theta_{11}^{C;A_3B_3}
\end{eqnarray}

\no$\underline{\theta^{13}}$

\begin{eqnarray}
M^{A_1A_2A_3}\Theta_{11}^{C;B_1B_2} &=&
\frac{1}{11} \biggl[-\frac{1}{36} \bigl(\Gamma^C\Gamma^{A_1A_2A_3}
- 12\eta^{CA_1}\Gamma^{A_2A_3}\bigr)\Theta_{13}^{B_1B_2} \nonumber \\
&+& \frac{1}{60} \bigl(5\eta^{CB_1}\Gamma^{B_2}\Gamma^{A_1}
- 18\eta^{CB_1}\eta^{B_2A_1} + 3\eta^{CA_1}\Gamma^{B_1B_2}
+ 3\eta^{A_1B_1}\Gamma^C\Gamma^{B_2}\bigr)\Theta_{13}^{A_2A_3} \nonumber \\
&+& \frac{1}{60} \bigl(2\Gamma^C\Gamma^{B_1}\Gamma^{A_1A_2}
- 8\eta^{CB_1}\Gamma^{A_1A_2} - 30\eta^{CA_1}\Gamma^{A_2}\Gamma^{B_1}
- 20\eta^{CA_1}\eta^{A_2B_1}\bigr)\Theta_{13}^{A_3B_2} \nonumber \\
&-& \frac{1}{36} \bigl(\Gamma^{A_1A_2A_3}\Gamma^{B_1}
+ 6\eta^{B_1A_1}\Gamma^{A_2A_3}\bigr)\Theta_{13}^{B_2C} \nonumber \\
&-& \frac{1}{60} \bigl(2\Gamma^{B_1B_2}\Gamma^{A_1A_2}
+ 30\eta^{A_1B_1}\Gamma^{B_2}\Gamma^{A_2}
- 80\eta^{A_1B_1}\eta^{A_2B_2}\bigr)\Theta_{13}^{A_3C}\biggr]
\end{eqnarray}

\begin{eqnarray}
M^{A_1A_2A_3}\Theta_{11}^{B_1 \ldots B_5} &=&
\frac{1}{72} \biggl(-\Gamma^{A_1A_2A_3}\Gamma^{B_1B_2B_3}
+ 6\eta^{B_1A_1}\Gamma^{A_2A_3}\Gamma^{B_2B_3}
+12\eta^{A_1B_1}\eta^{A_2B_2}\Gamma^{A_3}\Gamma^{B_3}\biggr)\Theta_{13}^{B_
4B_5} \nonumber \\
&+& \frac{1}{60} \biggl(\Gamma^{A_1A_2}\Gamma^{B_1 \ldots B_4}
+ 6\eta^{A_1B_1}\Gamma^{A_2}\Gamma^{B_2B_3B_4}
- 8\eta^{A_1B_1}\eta^{A_2B_2}\Gamma^{B_3B_4}\biggr)\Theta_{13}^{B_5A_3}
\nonumber \\
&+& \frac{2}{720} \biggl(\Gamma^{B_1 \ldots B_5}\Gamma^{A_1}
- 6\eta^{A_1B_1}\Gamma^{B_2 \ldots B_5}\biggr)\Theta_{13}^{A_2A_3}
\end{eqnarray}

\no$\underline{\theta^{15}}$

\be
M^{A_1A_2A_3}\Theta_{13}^{B_1B_2} =
-{1 \over 7 \times 720}\biggl(\Gamma^{A_1A_2A_3}\Gamma^{B_1B_2}
+ 6\eta^{B_1A_1}\Gamma^{A_2A_3}\Gamma^{B_2}
- 24\eta^{B_1A_1}\eta^{B_2A_2}\Gamma^{A_3}\biggr)\Theta
\ee

\section{Conclusions}

We have presented here in detail the irreducible tensors and spinor-tensors
contained in a scalar superfield of definite chirality, $\Phi(x, \theta^{(+)})$
in particular but the results for $\Phi(x, \theta^{(-)})$ are trivially
obtained making the changes explained in the introduction. The results for the
most basic products of these irreducible structures have also been presented
as a first step towards a full tensor calculus. The remaining products can be
derived by iteration of the formulae here and will appear elsewhere.

\vskip .5in

{\bf Acknowledgements.}

Part of the work of M. V. was done at the International School for Advanced
Studies, Trieste, Italy.

\ve
\setcounter{equation}{0}
\setcounter{section}{0}
\appendix
\renewcommand{\thesection}{Appendix \Alph{section}.}
\renewcommand{\theequation}{\Alph{section}.\arabic{equation}}

\section{Conventions and Bosonic Identities}

Our conventions are $\eta^{AB}=\eta_{AB}=diag(+- \ldots -)$,
$\epsilon^{01 \ldots 9}=\epsilon_{01 \ldots 9}=1$
and the Dirac algebra is

\be
\{\Gamma_A, \Gamma_B\} = 2 \eta_{AB} \;\;\; A,B=0,1, \ldots 9.
\ee
Our definition for $\Gamma_{(11)}$ is

\[\Gamma_{(11)} = \Gamma_0\Gamma_1 \ldots \Gamma_9 \]

\no which satisfies

\[ \Gamma_{(11)}^2 = I \qquad \Gamma_{(11)}^{\dagger} = \Gamma_{(11)} \]

Then $\theta^{(+)}=\Pi^{(+)}\theta=\frac{1}{2}(I + \Gamma_{(11)})\theta$
belongs to the
$[\frac{1}{2}\frac{1}{2}\frac{1}{2}\frac{1}{2}\frac{1}{2}]$
representation of SO(10) while
$\theta^{(-)}=\Pi^{(-)}\theta=\frac{1}{2}(I - \Gamma_{(11)})\theta$
belongs to
$[\frac{1}{2}\frac{1}{2}\frac{1}{2}\frac{1}{2}\frac{-1}{2}]$.

In 10 dimensions the Majorana and Weyl condition can be implemented
simultaneously and therefore our Majorana-Weyl spinors $\theta^{(\pm)}$
satisfy

\be
\bar\theta^{(\pm)} \Gamma_{A_1 \ldots A_n} \theta^{(\pm)}=0
\qquad \mbox{for  } n\ne 3,7
\ee

The only independent bilinear in $\theta^{(\pm)}$ is then
$M^{(\pm)}_{ABC} = \bar\theta^{(\pm)} \Gamma_{ABC}\theta^{(\pm)}$ since
we have the identity (2.8). Powers of this bilinear satisfy many identities,
implied by the basic Fierz one, that are used in the straightforward derivation
of the decomposition of the $\theta^6$ product in section III. Here is a list,

\[
M_{A_1A_2} \,\!^{C_1} M_{A_3} \,\!^{C_2C_3} =
-\frac{1}{3} M_{A_1A_2A_3} M^{C_1C_2C_3}  +
\delta_{A_1}^{C_1} M_{A_2A_3E} M^{C_2C_3E}
\]

\begin{eqnarray*}
M^{E[A_1A_2} M^{B_1B_2B_3]} &=&
\frac{2}{5} M^{EA_1A_2} M^{B_1B_2B_3} +
\frac{3}{5} M^{EB_1B_2} M^{B_3A_1A_2} \\
&-&\frac{3}{10} \eta^{EB_1} M^{A_1A_2D} M^{B_2B_3}\,\!_{D} -
\frac{3}{5} \eta^{A_1B_1} M^{A_2ED} M^{B_2B_3}\,\!_{D}
\end{eqnarray*}

\begin{eqnarray*}
\lefteqn{
M^{A_1C_1} \,\!_{D} M^{A_2C_2B_1} M^{B_2B_3D} =} \\
& &-\frac{1}{2} M_{D} \,\!^{C_1C_2} M^{A_1A_2B_1} M^{B_2B_3D}
+\frac{1}{2} M_{D} \,\!^{C_1C_2} M^{A_1A_2D} M^{B_1B_2B_3}
-\frac{1}{2} M^{B_1C_1C_2} M_{D} \,\!^{A_1A_2} M^{B_2B_3D} \\
& &-\frac{1}{2}
\eta^{A_1B_1} M^{A_2DE} M^{C_1C_2} \,\!_{D} M^{B_2B_3} \,\!_E
- \eta^{A_1C_1} M^{C_2DE} M^{A_2B_1} \,\!_{D} M^{B_2B_3} \,\!_E
- \frac{1}{2} \eta^{B_1C_1} M^{C_2DE} M^{A_1A_2} \,\!_{D} M^{B_2B_3} \,\!_E
\end{eqnarray*}

\begin{eqnarray*}
M^{A_3C_3B_3} M^{A_1A_2B_1} M^{C_1C_2B_2} &=&
\frac{1}{6} M^{B_1B_2B_3} M^{A_1A_2A_3} M^{C_1C_2C_3}
- \frac{1}{2} M^{A_3B_1B_2}  M^{B_3C_1C_2} M^{C_3A_1A_2} \\
&-&\frac{1}{2} \eta^{B_1C_1} M^{A_1A_2A_3} M^{B_2B_3D} M^{C_2C_3} \,\!_D
-\frac{1}{2} \eta^{A_1B_1} M^{A_2A_3} \,\!_D  M^{B_2B_3D} M^{C_1C_2C_3} \\
&-&\frac{1}{2} \eta^{A_1B_1} M^{A_2A_3} \,\!_D  M^{B_2B_3C_1} M^{C_2C_3D}
+\frac{1}{2} \eta^{A_1C_1} M^{A_2A_3} \,\!_D  M^{B_2B_3D} M^{C_2C_3B_1} \\
&+&\frac{1}{2} \eta^{A_1B_1}   M^{A_2A_3C_1} M^{B_2B_3} \,\!_D M^{C_2C_3D}
+ \eta^{B_1A_1} \eta^{B_2C_1} M^{B_3DE} M^{A_2A_3} \,\!_D M^{C_2C_3} \,\!_E
\end{eqnarray*}

\[
\epsilon^{AB_4 \ldots B_7CE_1 \ldots E_4} M^{DB_1B_2}
M_{DE_1E_2} M_{E_3E_4} \,\!^{B_3} = 0
\]

\[ M^{DE[A} M_{DF_1F_2} M_{F_3E} \,\!^{C]} = 0 \]

\[ M^{AB_1B_2} M^{B_3B_4B_5} M^{B_6B_7C} =
- \frac{2}{7 \times 5!} \epsilon^{B_1 \ldots B_7E_1E_2E_3}
M^{DAF} M_{DE_1E_2} M_{E_3F} \,\!^C \]

\begin{eqnarray*}
M^{AD_1D_2} M^{D_3D_4E} M^{C_1C_2} \,\!_E &=&
\frac{5}{6} M^{E[AD_1} M^{D_2D_3D_4]} M^{C_1C_2} \,\!_E \\
&+&\frac{5}{3} M^{[AD_1D_2} M^{C_1C_2]E} M^{D_3D_4} \,\!_E +
\frac{2}{3} \eta^{D_4C_1} M^{AE_1D_1} M^{D_2D_3E_2} M_{E_1E_2} \,\!^{C_2}
\end{eqnarray*}

\begin{eqnarray*}
\lefteqn{
\frac{1}{5!} \epsilon^{B_1 \ldots B_6D_1 \ldots D_4} M^{A} \,\!_{D_1D_2}
M_{D_3D_4E} M^{C_1C_2E} =} \\
& &\frac{2}{3} \frac{1}{5!} \epsilon^{B_1 \ldots B_6D_1D_2D_3C_1}
M^{AE_1} \,\!_{D_1} M_{D_2D_3}\,\!^{E_2} M_{E_1E_2} \,\!^{C_2} \\
& &+\frac{4}{3} \eta^{AB_1} M_E \,\!^{B_2B_3} M^{B_4B_5B_6} M^{C_1C_2E}
+\frac{2}{3} \eta^{C_1B_1} M_E \,\!^{B_2B_3} M^{B_4B_5B_6} M^{C_2AE}
\end{eqnarray*}

A curious identity in the $\theta^{10}$ sector that is easy to prove is

\[ \epsilon_{F_1F_2 \ldots F_{10}} M^{AF_1F_2} M^{BF_3F_4} M^{CF_5F_6}
M^{DF_7F_8} M^{EF_9F_{10}} = 0 \]

\no as it should be since no such symmetric object is allowed to exist.

Next we give a summary of how eq.(3.10) is derived directly from (2.12)
or (3.3-3.5). We start with

\begin{eqnarray*}
\lefteqn{M^{A_1A_2A_3}M^{B_1B_2B_3}M^{C_1C_2C_3} =} \\
& &= \biggl(- \frac{1}{4!} \epsilon^{B_1B_2B_3D_1 \ldots D_5 A_1A_2}
M^{A_3}\,\!_{D_1D_2} M_{D_3D_4D_5} +
\frac{3}{2} \eta^{A_1B_1} M^{A_2A_3} \,\!_E M^{B_2B_3E} \biggr)
M^{C_1C_2C_3} = \\
& &= - \frac{1}{4!} \epsilon^{B_1B_2B_3D_1 \ldots D_5 A_1A_2}
M^{A_3}\,\!_{D_1D_2} \biggl(
- \frac{1}{4!} \epsilon_{D_3D_4D_5}\,\!^{E_1 \ldots E_5 C_1C_2}
M^{C_3}\,\!_{E_1E_2} M_{E_3E_4E_5} +
\frac{3}{2} \delta^{C_1}_{D_3} M_{D_4D_5E} M^{C_2C_3E} \biggr) \\
& &+ \frac{3}{2} \eta^{A_1B_1} M^{A_2A_3} \,\!_E M^{B_2B_3E}
M^{C_1C_2C_3}
\end{eqnarray*}

Expanding the product of the Levi-Civita symbols and using heavily the
identities above, one gets after a lot algebra

\[M^{A_1A_2A_3}M^{B_1B_2B_3}M^{C_1C_2C_3} =
\frac{9}{8} M^{A_3C_1C_2}M^{B_3A_1A_2}M^{C_3B_1B_2} +
I^{A_1A_2A_3B_1B_2B_3C_1C_2C_3}\]

\no with

\begin{eqnarray*}
\lefteqn{I^{A_1A_2A_3B_1B_2B_3C_1C_2C_3} = }\\
& & = \frac{3}{8} \biggl\{
6 \eta^{B_1C_1} M^{A_1A_2A_3}M^{B_2B_3D}M^{C_2C_3}\,\!_D
+ 4 \eta^{A_1C_1} M^{A_2A_3D}M^{B_1B_2B_3}M^{C_2C_3}\,\!_D \\
& &+ \frac{9}{2} \eta^{A_1B_1} M^{A_2A_3D}M^{B_2B_3}\,\!_DM^{C_1C_2C_3}
- 6 \eta^{A_1C_1} M^{A_2A_3B_1}M^{B_2B_3D} M^{C_2C_3}\,\!_D \\
& &+ \frac{3}{2} \eta^{A_1B_1} \biggl(
M^{A_2A_3C_1}M^{B_2B_3D}M^{C_2C_3}\,\!_D
- M^{A_2A_3D}M^{B_2B_3C_1}M^{C_2C_3}\,\!_D \biggr) \\
& &- \frac{9}{2} \eta^{B_1C_1} M^{A_1A_2D}M^{B_2B_3}\,\!_D M^{A_3C_2C_3}
- 3 \eta^{A_1B_1}\eta^{A_2C_1} M^{A_3}\,\!_{DE}M^{B_2B_3D}M^{C_2C_3E} \\
& &- 6 \eta^{A_1C_1}\eta^{B_1C_2} M^{C_3}\,\!_{DE}M^{A_2A_3D}M^{B_2B_3E}
- 6 \eta^{A_1B_1}\eta^{B_2C_1} M^{B_3}\,\!_{DE}M^{A_2A_3D}M^{C_2C_3E} \\
& &- 3\biggl( \eta^{A_1C_1}\eta^{A_2C_2} M^{B_1B_2D}M^{B_3A_3E}
+ \eta^{B_1C_1}\eta^{B_2C_2} M^{A_1A_2D}M^{A_3B_3E} \biggr)M^{C_3}\,\!_{DE}
\biggr\} \\
& &- \epsilon^{B_1B_2B_3D_1 \ldots D_4C_1A_1A_2}
M^{A_3}\,\!_{D_1D_2} M_{D_3D_4E} M^{C_2C_3E}
\end{eqnarray*}

Iterating this equation, we arrive at

\begin{eqnarray*}
\lefteqn{M^{A_3C_1C_2}M^{B_3A_1A_2}M^{C_3B_1B_2} -
I^{C_1C_2A_3B_1B_2C_3A_1A_2B_3} =
\frac{9}{8} M^{\mbox{}^{\scriptstyle [A_3} [A_1A_2} M^{B_3][B_1B_2}
M^{C_3]\mbox{}^{\scriptstyle C_1C_2]}} = }\\
& &= \frac{5}{24} M^{A_1A_2A_3}M^{B_1B_2B_3}M^{C_1C_2C_3} \\
& &- \frac{1}{6} \biggl(M^{A_3C_1C_2}M^{B_3A_1A_2}M^{C_3B_1B_2}
+ \frac{1}{2} M^{A_3B_1B_2}M^{B_3C_1C_2}M^{C_3A_1A_2} \biggr) \\
& & + II^{A_1A_2A_3B_1B_2B_3C_1C_2C_3}
\end{eqnarray*}

\no with

\begin{eqnarray*}
\lefteqn{II^{A_1A_2A_3B_1B_2B_3C_1C_2C_3} =} \\
& &- \frac{1}{4} \biggl(
\eta^{B_1C_1} M^{A_1A_2A_3}M^{B_2B_3D}M^{C_2C_3}\,\!_D
+ \eta^{A_1C_1} M^{A_2A_3D}M^{B_1B_2B_3}M^{C_2C_3}\,\!_D \\
& &+ \eta^{A_1B_1} M^{A_2A_3D}M^{B_2B_3D}M^{C_1C_2C_3} \biggr) \\
& &+ \frac{1}{12} \biggl(
\eta^{A_1C_1} M^{A_2A_3B_1}M^{B_2B_3D}M^{C_2C_3}\,\!_D
+ \eta^{B_1C_1} M^{A_1A_2}\,\!_D M^{B_2B_3D}M^{C_2C_3A_3} \biggr) \\
& &+ \frac{1}{6} \biggl(
\eta^{A_1B_1}\eta^{A_2C_1} M^{A_3}\,\!_{DE}M^{B_2B_3D}M^{C_2C_3E}
+ \eta^{A_1C_1}\eta^{B_1C_2} M^{C_3}\,\!_{DE}M^{A_2A_3D}M^{B_2B_3E} \\
& &+ \eta^{A_1B_1}\eta^{B_2C_1} M^{B_3}\,\!_{DE}M^{A_2A_3D}M^{C_2C_3E} \biggr)
\end{eqnarray*}

Applying the (normalized) operator ${\cal S}(A,B,C)$ that fully symmetrizes
upon interchange of the letters $A,B,C$, to the equations we have just
obtained, we get a system of two equations with solution

\begin{eqnarray*}
M^{A_1A_2A_3}M^{B_1B_2B_3}M^{C_1C_2C_3} &=&
\frac{16}{65} {\cal S}(A, B, C) \biggl[
5 I^{A_1A_2A_3B_1B_2B_3C_1C_2C_3}  \\
&+& \frac{9}{2} \biggl(
I^{C_1C_2A_3B_1B_2C_3A_1A_2B_3} +
II^{A_1A_2A_3B_1B_2B_3C_1C_2C_3} \biggr) \biggr]
\end{eqnarray*}

Let us now proceed to prove the duality properties of the tensors
${\cal M}_{12}^{A;B_1 \ldots B_5}$ and
${\cal M}_{10}^{CD;B_1 \ldots B_5}$. From (4.34) and (\ref{friend}) we
can also write

\be
{\cal M}_{12}^{B;A_1 \ldots A_5}
= \frac{1}{2} M^B\,\!_{D_1}\,\!^{D_2} M^F\,\!_{D_2}\,\!^{D_3}
M^{A_1}\,\!_{D_3}\,\!^{D_4} M^{D_1}\,\!_{D_4}\,\!^{D_5}
M^{A_2A_3}\,\!_{D_5} M_F\,\!^{A_4A_5}
\ee

But:

\begin{eqnarray*}
\lefteqn{M^B\,\!_{D_1}\,\!^{D_2} M_{FD_2}\,\!^{D_3}
M^{A_1}\,\!_{D_3}\,\!^{D_4} M^{D_1}\,\!_{D_4}\,\!^{D_5}
M^{[A_2A_3}\,\!_{D_5} M^{FA_4A_5]} =} \\
& &=M^B\,\!_{D_1}\,\!^{D_2} M_{FD_2}\,\!^{D_3}
M^{A_1}\,\!_{D_3}\,\!^{D_4} M^{D_1}\,\!_{D_4}\,\!^{D_5}
\frac{1}{5} \left(3 M^{A_2A_3}\,\!_{D_5} M^{FA_4A_5}
+ 2M^{FA_2}\,\!_{D_5} M^{A_3A_4A_5} \right) \\
& &=\frac{3}{5}
M^B\,\!_{D_1}\,\!^{D_2} M_{FD_2}\,\!^{D_3}
M^{A_1}\,\!_{D_3}\,\!^{D_4} M^{D_1}\,\!_{D_4}\,\!^{D_5}
M^{A_2A_3}\,\!_{D_5} M^{FA_4A_5},
\end{eqnarray*}

\no so

\begin{eqnarray*}
{\cal M}_{12}^{B;A_1 \ldots A_5} &=&
\frac{5}{6} M^B\,\!_{D_1}\,\!^{D_2} M_{FD_2}\,\!^{D_3}
M^{A_1}\,\!_{D_3}\,\!^{D_4} M^{D_1}\,\!_{D_4}\,\!^{D_5}
M^{[A_2A_3}\,\!_{D_5} M^{FA_4A_5]} \\
&=&-\frac{5}{6} \frac{1}{5!} M^B\,\!_{D_1}\,\!^{D_2} M_{FD_2}\,\!^{D_3}
M^{A_1}\,\!_{D_3}\,\!^{D_4} M^{D_1}\,\!_{D_4}\,\!^{D_5}
\epsilon^{FA_2 \ldots A_5E_1 \ldots E_5} M_{D_5E_1E_2} M_{E_3E_4E_5}
\end{eqnarray*}

Now we have to ``rotate" indices; that is, from the identity:

\be
M_{FD_2}\,\!^{D_3}M^{D_1}\,\!_{D_4}\,\!^{D_5}M_{D_3}\,\!^{D_4[A_1}
\epsilon^{FA_2 \ldots A_5E_1 \ldots E_5]} M_{D_5E_1E_2} M_{E_3E_4E_5}=0
\ee

\no we see that

\begin{eqnarray*}
\lefteqn{
M_{FD_2}\,\!^{D_3}M^{D_1}\,\!_{D_4}\,\!^{D_5}M_{D_5E_1E_2} M_{E_3E_4E_5}
\biggl[5M_{D_3}\,\!^{D_4A_1}
\epsilon^{FA_2 \ldots A_5E_1 \ldots E_5} } \\
& &-M_{D_3}\,\!^{D_4F}
\epsilon^{A_1 \ldots A_5E_1 \ldots E_5}
-2M_{D_3}\,\!^{D_4E_1}
\epsilon^{FA_1 \ldots A_5E_2 \ldots E_5}
-3M_{D_3}\,\!^{D_4E_3}
\epsilon^{FA_1 \ldots A_5E_1E_2E_4E_5}
\biggr] = 0,
\end{eqnarray*}

\no the second and third term vanish identically because of (2.3) and
(\ref{nueva}) respectively, and we obtain

\begin{eqnarray*}
\lefteqn{
M_{FD_2}\,\!^{D_3}M^{D_1}\,\!_{D_4}\,\!^{D_5}M_{D_5E_1E_2} M_{E_3E_4E_5}
M_{D_3}\,\!^{D_4A_1}
\epsilon^{FA_2 \ldots A_5E_1 \ldots E_5} =} \\
& &=\frac{3}{5}
M_{FD_2}\,\!^{D_3}M^{D_1}\,\!_{D_4}\,\!^{D_5}M_{D_5E_1E_2} M_{HE_3E_4}
M_{D_3}\,\!^{D_4H}
\epsilon^{FA_1 \ldots A_5E_1 \ldots E_4} .
\end{eqnarray*}

Therefore

\begin{eqnarray*}
{\cal M}_{12}^{B;A_1 \ldots A_5} &=&\frac{1}{2 \times 5!}
\epsilon^{A_1 \ldots A_5FE_1 \ldots E_4}
M^B\,\!_{D_1}\,\!^{D_2} M^{D_1}\,\!_{D_4}\,\!^{D_5}
M_{FD_2}\,\!^{D_3}M_{D_3}\,\!^{D_4H}
M_{D_5E_1E_2} M_{HE_3E_4} \\
&=&-\frac{1}{2 \times 5!}
\epsilon^{A_1 \ldots A_5FE_1 \ldots E_4}
M^B\,\!_{D_1}\,\!^{D_2} M^{D_1}\,\!_{D_4}\,\!^{D_5}
M^{D_4}\,\!_{F}\,\!^{D_3}M_{D_3D_2}\,\!^{H}
M_{D_5E_1E_2} M_{HE_3E_4} \\
&=&\frac{1}{2 \times 5!}
\epsilon^{A_1 \ldots A_5E_1 \ldots E_5}
M^{BD_2}\,\!_{D_1} M^{D_5D_1}\,\!_{D_4}
M_{E_1}\,\!^{D_4D_3}M_{D_2D_3}\,\!^{H}
M_{D_5E_2E_3} M_{HE_4E_5} \\
&=&\frac{1}{5!}
\epsilon^{A_1 \ldots A_5E_1 \ldots E_5}
{\cal M}_{12}^{B;}\,\!_{A_1 \ldots A_5},
\end{eqnarray*}

\no the desired result. Notice the opposite sign with respect to the
$\theta^4$ piece, whose duality was explicitly used. For
${\cal M}_{10}^{CD;B_1 \ldots B_5}$ the derivation proceeds similarly and
again one obtains a result opposite to the $\theta^4$ one.

\ve
\section{Fermionic Identities}
\setcounter{equation}{0}

In this Appendix we list identities involving some products of powers of
$M^{ABC}$ with the spinor-tensors.

\be
M_{E_1E_2}\,\!^C\Theta_5^{A_1A_2A_3E_1E_2} =
\frac{3}{5} \Theta_7^{C;A_1A_2A_3}
\ee

\be
{\cal M}_4^{EB;A_1A_2}\Theta_3\,\!^C\,\!_E =
\frac{1}{2} \biggl(\hat{\Theta}_7^{A_1A_2;BC}
- \hat{\Theta}_7^{BA_1;A_2C}\biggr)
+ \frac{1}{4}\Gamma^{A_1}\Theta_7^{A_2BC}
\ee

\begin{eqnarray}
\lefteqn{M^{CE_1E_2}M^{A_1A_2}\,\!_{E_1}M^{B_1B_2}\,\!_{E_2}\theta =}
\nonumber \\
& &=\frac{1}{28} \biggl(\Gamma^{C}\hat{\Theta}_7^{A_1A_2;B_1B_2}
+ 2\Gamma^{A_1}\hat{\Theta}_7^{A_2B_1;B_2C}
- 2\Gamma^{A_1}\hat{\Theta}_7^{A_2C;B_1B_2}
- \Gamma^{A_1}\Gamma^{B_1}\Theta_7^{A_2B_2C} \nonumber \\
& &-\Gamma^{C}\hat{\Theta}_7^{B_1B_2;A_1A_2}
- 2\Gamma^{B_1}\hat{\Theta}_7^{B_2A_1;A_2C}
+ 2\Gamma^{B_1}\hat{\Theta}_7^{B_2C;A_1A_2}
+ \Gamma^{B_1}\Gamma^{A_1}\Theta_7^{B_2A_2C}\biggr)
\end{eqnarray}

\be
\hat{\Theta}_7^{BA_1;A_2C} = \frac{3}{2} \Theta_7^{(B;C)A_1A_2}
\ee

\be
\Gamma_{E}M^{EBF}\hat{\Theta}_7\,\!_F\,\!^{C;A_1A_2} =
M^{A_1A_2E}\Theta_7^{BC}\,\!_{E}
\ee

\be
\Theta_9^{ABC} =
\frac{1}{24} \Gamma_{E_1E_2E_3}M^{E_1E_2E_3}\Theta_7^{ABC}
\ee

\be
M^{AD}\,\!_{E}M^{B}\,\!_{DF}\Theta_7^{CEF} = \hat{\Theta}_{11}^{C;AB}
\ee

\be
M_{E_1E_2}\,\!^A\Theta_9^{B;CE_1E_2} =
\frac{2}{3} \hat{\Theta}_{11}^{B;AC}
\ee

\be
\Theta_9^{(C;D)A_1A_2} = \frac{2}{3} M^{EA_1A_2}\Theta_7^{CD}\,\!_E
+ \frac{1}{3} \Gamma^{A_1}\Theta_9^{A_2CD}
\ee

\begin{eqnarray}
\lefteqn{M_{E_1}\,\!^{AE_2}M_{E_2}\,\!^{BE_3}M_{E_3}\,\!^{CE_4}
M_{E_4}\,\!^{DE_5} \Theta_3\,\!_{E_5}\,\!^{E_1} =} \nonumber \\
& &=\frac{1}{42} \biggl(2\Gamma^A\hat{\Theta}_{11}^{B;CD}
+ \Gamma^A\hat{\Theta}_{11}^{C;BD} - 4\Gamma^B\hat{\Theta}_{11}^{A;CD}
+ \Gamma^B\hat{\Theta}_{11}^{C;AD} \nonumber \\
& &- 4\Gamma^C\hat{\Theta}_{11}^{A;BD}
- 5\Gamma^C\hat{\Theta}_{11}^{B;AD} - \Gamma^D\hat{\Theta}_{11}^{B;AC}
- 2\Gamma^D\hat{\Theta}_{11}^{C;AB}\biggr)
\end{eqnarray}

\be
\Gamma_{E}M^{EA}\,\!_{F}\hat{\Theta}_{11}^{F;BC} =
-\frac{1}{3} \Gamma^{(B}\Theta_{13}^{C)A}
\ee

\be
\Gamma_{E}M^{EA}\,\!_{F}\hat{\Theta}_{11}^{B;CF} =
\frac{1}{2} \Gamma^A\Theta_{11}^{BC} +  \frac{1}{3} \Gamma^B\Theta_{11}^{CA}
+ \frac{1}{6}\Gamma^C\Theta_{11}^{AB}
\ee

\ve
\section{Young Projector Method}
\setcounter{equation}{0}

Let us consider a Young diagram $R$ with $n$ rows having $m_i$ boxes in the
$i^{th}$ row $(m_1 \geq m_2 \geq \ldots \geq m_n)$ and having $\lambda_j$
boxes in the $j^{th}$ column
$(n=\lambda_1 \geq \lambda_2 \geq \ldots \geq \lambda_{m_1})$. The Young
projector corresponding to a particular ($R_I$)
{\it standard tableau} \cite{Bacry}
is given by

\be
Y\left(R_I\right) = \alpha(R) {\cal Q} {\cal P}
\label{B1}
\ee
\[{\cal Q} = \prod_{i=1}^{m_1} Q_i \;\;\; {\cal P} = \prod_{j=1}^{n} P_j \]

\no where $P_j$ is the (normalized) operator that fully symmetrizes over the
entries of the $j^{th}$ row and $Q_i$ is the (normalized) one that fully
antisymmetrizes over the entries of the $i^{th}$ column. For operators so
normalized, the normalization factor $\alpha$ needed for $Y$ to be
idempotent $Y^2=Y$, is

\be
\alpha(R) = \frac{dim(R)}{m!} \left(\prod_{j=1}^{n} m_j! \right)
\left( \prod_{i=1}^{m_1} \lambda_i! \right)
\label{B2}
\ee

\no where $m=\sum_{j=1}^n m_j = \sum_{i=1}^{\lambda_i}$ is the total number of
boxes in the Young diagram and $dim(R)$ is the dimension of the irreducible
representation of the symmetric group ${\cal S}_m$ corresponding to the
diagram $R$ \cite{Barut}.
The products of factorials in (\ref{B2}) appear because we considered
normalized $Q_i$ and $P_j$ in (\ref{B1}) ($Q_i^2 = Q_i, P_j^2 = P_j$).

There are 14 standard tableaux associated with the diagram
\setlength{\unitlength}{.5pt}
$\begin{picture}(20, 50)
\multiput(0, 30)(0, -10){5}{\framebox(10,10){ }}
\multiput(10, 30)(0, -10){2}{\framebox(10,10){ }}
\end{picture}$ ,
however, due to identity (\ref{friend}) many of them do not contribute.
The tableaux that give non-vanishing results are

\be
\setlength{\unitlength}{1pt}
\begin{picture}(20, 50)
\put(0, 20){1}
\put(0, 10){2}
\put(0, 0){3}
\put(0, -10){6}
\put(0, -20){7}
\put(10, 20){4}
\put(10, 10){5}
\end{picture}
\;\;\;
\begin{picture}(20, 50)
\put(0, 20){1}
\put(0, 10){2}
\put(0, 0){3}
\put(0, -10){5}
\put(0, -20){7}
\put(10, 20){4}
\put(10, 10){6}
\end{picture}
\;\;\;
\begin{picture}(20, 50)
\put(0, 20){1}
\put(0, 10){2}
\put(0, 0){3}
\put(0, -10){5}
\put(0, -20){6}
\put(10, 20){4}
\put(10, 10){7}
\end{picture}
\;\;\;
\begin{picture}(20, 50)
\put(0, 20){1}
\put(0, 10){2}
\put(0, 0){3}
\put(0, -10){4}
\put(0, -20){7}
\put(10, 20){5}
\put(10, 10){6}
\end{picture}
\;\;\;
\begin{picture}(20, 50)
\put(0, 20){1}
\put(0, 10){2}
\put(0, 0){3}
\put(0, -10){4}
\put(0, -20){6}
\put(10, 20){5}
\put(10, 10){7}
\end{picture}
\;\;\;
\begin{picture}(20, 50)
\put(0, 20){1}
\put(0, 10){2}
\put(0, 0){3}
\put(0, -10){4}
\put(0, -20){5}
\put(10, 20){6}
\put(10, 10){7}
\end{picture}
\label{TB1}
\ee
\vskip1cm

\no and the results for all the tableaux can be inferred from the first two

\begin{eqnarray}
\lefteqn{Y\left(
\setlength{\unitlength}{1pt}
\begin{picture}(20, 30)
\put(0, 20){1}
\put(0, 10){2}
\put(0, 0){3}
\put(0, -10){6}
\put(0, -20){7}
\put(10, 20){4}
\put(10, 10){5}
\end{picture}
\right) M^{A_1A_2A_3} M^{B_1B_2D} M^{C_1C_2} \,\!_D =}
\nonumber \\
& & = \frac{\alpha}{4} \bigl( M^{[A_1A_2A_3} M^{C_1C_2]} \,\!_D M^{B_1B_2D}
+ M^{[B_1A_2A_3} M^{C_1C_2]} \,\!_D M^{A_1B_2D}
\nonumber \\
& &+ M^{[A_1B_2A_3} M^{C_1C_2]} \,\!_D M^{B_1A_2D}
+ M^{[B_1B_2A_3} M^{C_1C_2]} \,\!_D M^{A_1A_2D} \bigr)
\label{NL1}
\end{eqnarray}

\begin{eqnarray}
\lefteqn{Y\left(
\setlength{\unitlength}{1pt}
\begin{picture}(20, 30)
\put(0, 20){1}
\put(0, 10){2}
\put(0, 0){3}
\put(0, -10){5}
\put(0, -20){7}
\put(10, 20){4}
\put(10, 10){6}
\end{picture}
\right) M^{A_1A_2A_3} M^{B_1B_2D} M^{C_1C_2} \,\!_D =}
\nonumber \\
& & = \frac{\alpha}{8} \bigl(M^{[A_1A_2A_3} M^{B_2C_2]} \,\!_D M^{C_1B_1D}
+ M^{[B_1A_2A_3} M^{B_2C_2]} \,\!_D M^{A_1C_1D}
\nonumber \\
& &+ M^{[A_1C_1A_3} M^{B_2C_2]} \,\!_D M^{B_1A_2D}
+ M^{[B_1C_1A_3} M^{B_2C_2]} \,\!_D M^{A_1A_2D} \bigr) ,
\label{NL2}
\end{eqnarray}

\no the letter convention has been momentarily suspended in (\ref{NL1}) and
(\ref{NL2}).

So, to obtain the total projection corresponding to the diagram
\setlength{\unitlength}{.5pt}
\begin{picture}(20, 50)
\multiput(0, 20)(0, -10){5}{\framebox(10,10){ }}
\multiput(10, 20)(0, -10){2}{\framebox(10,10){ }}
\end{picture}
we add the contributions of all the standard tableaux in (\ref{TB1})

\begin{eqnarray}
\lefteqn{Y\left(
\setlength{\unitlength}{.5pt}
\begin{picture}(20, 40)
\multiput(0, 20)(0, -10){5}{\framebox(10,10){ }}
\multiput(10, 20)(0, -10){2}{\framebox(10,10){ }}
\end{picture}
\right) M^{A_1A_2A_3} M^{B_1B_2D} M^{C_1C_2} \,\!_D =}
\nonumber \\
& & = \frac{\alpha}{4} \bigl(
M^{[A_1A_2A_3} M^{B_1B_2]} \,\!_D M^{C_1C_2D} +
M^{[A_1A_2A_3} M^{C_1C_2]} \,\!_D M^{B_1B_2D}
\nonumber \\
& &+ 2 M^{[A_1A_2A_3} M^{B_1C_1]} \,\!_D M^{B_2C_2D} \bigr)
\end{eqnarray}

\[
\alpha = \frac{dim\left(
\setlength{\unitlength}{.5pt}
\begin{picture}(20, 40)
\multiput(0, 20)(0, -10){5}{\framebox(10,10){ }}
\multiput(10, 20)(0, -10){2}{\framebox(10,10){ }}
\end{picture}
\right)}{7!} (2!2!)(5!2!) = \frac{8}{3}
\]

A comment is in order here. In projecting an arbitrary tensor one obtains
a {\it different} irreducible representation for {\it each} standard
tableau \cite{Bacry}. The same is not true here, of course, because of the
nilpotency of the $\theta$-tensors. Each irreducible representation appears
only once at each level in Table 1. The number of degrees of freedom are
dramatically reduced by the nilpotency of these structures and that is why
the problem becomes manageable. For instance, the product $M^{A_1A_2A_3}
M^{B_1B_2B_3}$ instead of having
$\left(\begin{array}{c}{10}\\{3}\end{array}\right)
\times \left(\begin{array}{c}{10}\\{3}\end{array}
\right) = 120^2 = 14400$
degrees of freedom, it has only
$\left(\begin{array}{c}{16}\\{4}\end{array}\right)
= 770 + 1050 = 1820$. But
doing the counting explicitly by subtracting the number of independent
constraints implied by the conditions on the irreducible pieces and
otherwise derivable identities, can be an extremely painful task. However,
one does not need to dwell into all that detail, fortunately, but rather
proceed to add all the projectors for the different standard tableaux
corresponding to a Young diagram in order to consistently extract the
{\it unique} representation involved in all the cases.

\end{document}